\documentclass[pra,aps,twocolumn,notitlepage,superscriptaddress,showpacs,nofootinbib]{revtex4-1}
\usepackage[colorlinks]{hyperref}
\hypersetup{
	colorlinks = true,
	citecolor = {blue},
	linkcolor = {blue},
	urlcolor = {blue},	
}
\usepackage{verbatim}
\usepackage{dsfont}
\usepackage{graphicx} 
\usepackage{bm}      
\usepackage[usenames]{color}
\usepackage{color}
\usepackage{amsmath} 
\usepackage[10pt]{moresize}
\usepackage[latin1]{inputenc}
\usepackage{latexsym}
\usepackage{amssymb}   
\usepackage{booktabs} 
\usepackage{multirow}  
\usepackage{subfigure} 
\usepackage{dcolumn}
\usepackage{mathrsfs}

\newcommand{\RNum}[1]{\uppercase\expandafter{\romannumeral #1\relax}}

\newcommand{\ket}[1]{\ensuremath{|#1\rangle}}

\newcommand{\bra}[1]{\ensuremath{\langle#1|}}

\usepackage{ulem}

\newcommand{\beq}{\begin{equation}}
\newcommand{\eeq}{\end{equation}}
\newcommand{\bqa}{\begin{eqnarray}}
\newcommand{\eqa}{\end{eqnarray}}

\newcommand{\Tr}{\textrm{Tr}}

\newcommand{\forget}[1]{}
\usepackage{graphicx}

\graphicspath{{FIG/}}

\begin{document}
\preprint{APS/123-QED}
\title{Error-Disturbance Trade-off in Sequential Quantum Measurements}

\author{Ya-Li Mao}
\affiliation{Department of Modern Physics and National Laboratory for Physical Sciences at Microscale, Shanghai Branch, University of Science and Technology of China, Hefei, Anhui 230026, China}
\affiliation{CAS Center for Excellence and Synergetic Innovation Center in Quantum Information and Quantum Physics, Shanghai Branch,  University of Science and Technology of China, Hefei, Anhui 230026, China}

\author{Zhi-Hao Ma}
\affiliation{School of Mathematical Sciences, Shanghai Jiao Tong University, Shanghai 200240,  China}

\author{Rui-Bo Jin}
\affiliation{Hubei Key Laboratory of Optical Information and Pattern Recognition, Wuhan Institute of Technology, Wuhan 430205, China}

\author{Qi-Chao Sun}
\affiliation{Department of Modern Physics and National Laboratory for Physical Sciences at Microscale, Shanghai Branch, University of Science and Technology of China, Hefei, Anhui 230026, China}
\affiliation{CAS Center for Excellence and Synergetic Innovation Center in Quantum Information and Quantum Physics, Shanghai Branch,  University of Science and Technology of China, Hefei, Anhui 230026, China}

\author{Shao-Ming Fei}
\affiliation{School of Mathematical Sciences, Capital Normal University, Beijing 100048, China}
\affiliation{Max-Planck-Institute for Mathematics in the Sciences, 04103 Leipzig, Germany}

\author{Qiang Zhang}
\affiliation{Department of Modern Physics and National Laboratory for Physical Sciences at Microscale, Shanghai Branch, University of Science and Technology of China, Hefei, Anhui 230026, China}
\affiliation{CAS Center for Excellence and Synergetic Innovation Center in Quantum Information and Quantum Physics, Shanghai Branch,  University of Science and Technology of China, Hefei, Anhui 230026, China}

\author{Jingyun Fan}
\affiliation{Department of Modern Physics and National Laboratory for Physical Sciences at Microscale, Shanghai Branch, University of Science and Technology of China, Hefei, Anhui 230026, China}
\affiliation{CAS Center for Excellence and Synergetic Innovation Center in Quantum Information and Quantum Physics, Shanghai Branch,  University of Science and Technology of China, Hefei, Anhui 230026, China}

\author{Jian-Wei Pan}
\affiliation{Department of Modern Physics and National Laboratory for Physical Sciences at Microscale, Shanghai Branch, University of Science and Technology of China, Hefei, Anhui 230026, China}
\affiliation{CAS Center for Excellence and Synergetic Innovation Center in Quantum Information and Quantum Physics, Shanghai Branch,  University of Science and Technology of China, Hefei, Anhui 230026, China}

\begin{abstract}
We derive a state-dependent error-disturbance trade-off based on a statistical distance in the sequential measurements of a pair of noncommutative observables and experimentally verify the relation with a photonic qubit system. We anticipate that this Letter may further stimulate the study on the quantum uncertainty principle and related applications in quantum measurements.
\end{abstract}

\maketitle

{\it Introduction}.---Uncertainty is an essential feature of quantum mechanics, which underlies the quantum measurement and the emerging quantum information science \cite{RevModPhys.86.1261,RevModPhys.89.015002} and is best reflected in the joint measurements of a pair of noncommutative observables $\hat{A}$ and $\hat{B}$. When measured separately, their measurement precisions, denoted by standard deviations $\sigma (\hat{A})$ and $\sigma(\hat{B})$, are jointly constrained by the celebrated Heisenberg-Robertson relation, $\sigma (\hat{A})\sigma(\hat{B}) \geq \frac{1}{2}\langle[\hat{A},\hat{B}]\rangle$~\cite{Heisenberg1927,Kennard1927,robertson1929uncertainty}. For sequential (or joint) measurements that are an important aspect of the quantum uncertainty principle, a simple and intuitive trade-off between the error $\epsilon(\hat{A})$ and the disturbance $\eta(\hat{B})$ has remained a long-sought goal~\cite{Scully1991nature,Storey1994nature,Wiseman1995nature}.
Heisenberg's intuition, $\epsilon(\hat{A}) \eta(\hat{B}) \geq \frac{1}{2}|\langle[\hat{A},\hat{B}]\rangle|$ \cite{Heisenberg1927}, was negated in the experiments \cite{erhart2012experimental,PhysRevLett.109.100404,baek2013experimental,PhysRevA.88.022110,PhysRevA.88.022110,PhysRevLett.112.020402,PhysRevLett.112.020401}. Ozawa recently showed that the error and disturbance quantified by the root-mean-square (rms) deviations, $\epsilon(\hat{A}) =  \langle( \hat{N}_{\hat{A}}-\hat{A} \otimes \hat{I})^2 \rangle^{(1/2)}$ and $\eta(\hat{B}) = \langle(\hat{N}_{\hat{B}}-\hat{B} \otimes \hat{I})^2 \rangle^{(1/2)}$, satisfy the following relation \cite{Ozawa03,ozawa2004uncertainty,OZAWA2004367},
\begin{equation}\label{ozawa}
\epsilon(\hat{A}) \eta(\hat{B}) + \epsilon(\hat{A}) \sigma(\hat{B}) + \sigma(\hat{A}) \eta(\hat{B})\geq \frac{1}{2}|\langle[\hat{A},\hat{B}]\rangle|,
\end{equation}
where $\hat{N}_{\hat{A}}=\hat{U}^\dagger(\hat{I} \otimes \hat{M})\hat{U} $ and $\hat{N}_{\hat{B}}= \hat{U}^\dagger(\hat{B} \otimes \hat{I})\hat{U}$ are noise operators, and the unitary operator $\hat{U}$ couples the system state $\rho$ to the meter state $\mu$.
$\epsilon(\hat{A})$ [$\eta(\hat{B})$] is extracted as the rms difference between the value of $\hat{A}$ ($\hat{B}$) measured with the initial system state $\rho$ and the value of $\hat{M}$ ($\hat{B}$) measured with the coupled state $\hat{U}^+(\rho\otimes\mu)\hat{U}$.
$\epsilon(\hat{A})$ and $\eta(\hat{B})$ are analogs of the classical rms values. Relation (\ref{ozawa}) was soon verified in a number of experiments \cite{erhart2012experimental,PhysRevLett.109.100404,baek2013experimental,PhysRevA.88.022110}, which thereafter has stimulated a great deal of interest in the investigation of joint measurability of two observables with fruitful outcomes \cite{branciard2013error,RevModPhys.89.015002,PhysRevLett.112.020402,PhysRevLett.110.120403,PhysRevLett.110.220402,Werner1,PhysRevA.89.022124,PhysRevLett.112.050401,Werner2,Werner3,PhysRevA.90.042113,PhysRevA.89.022106,PhysRevA.89.052108,PhysRevLett.112.020401,RevModPhys.86.1261,PhysRevLett.115.030401,PhysRevLett.114.070402,PhysRevLett.117.140402,EXPW1,EXPW2,PhysRevA.96.022137,Barchielli2018}.

Quantum mechanics is fundamentally probabilistic. A quantum measurement gives rise to a probability distribution associated with the eigenvalues of the measurement observable. While the rms (or a similar value comparison) approach uses the information of both values and probability distributions of observables, studying the joint measurability from the information theoretic perspective, which characterizes error and disturbance only based on probability distributions in the format of relative entropy, statistical distances, etc., has also been explored in depth \cite{Werner1,Werner2,Werner3,PhysRevLett.112.050401,PhysRevLett.115.030401,EXPW1,EXPW2,PhysRevA.96.022137,Barchielli2018}.
Inspired by this remarkable progress, we present in this Letter a new study on the error-disturbance trade-off. By adopting the probability distance used in characterizing the incompatibility of the two measurements \cite{Werner1,Werner3}, we show that the error and disturbance trade-off relation in the sequential measurements are naturally constrained by a triangle inequality. This relation satisfies the natural and significant requirements for operational error and disturbance~\cite{PhysRevA.89.052108}, is free of the shortcomings of relation (\ref{ozawa}) and is state dependent.

{\it Main results}.---Consider a given quantum state $\rho$ and two observables, $\hat{A}=\sum\limits_{i}a_i\hat{A}_{i}$ and $\hat{B}=\sum\limits_{i}b_i\hat{B}_{i}$, where $\{\hat{A}_i\}$ and $\{\hat{B}_i\}$ ($\{a_i\}$ and $\{b_i\}$) are the projective measurements with respect to the eigenvectors (eigenvalues) of $\hat{A}$ and $\hat{B}$, respectively. The measurement $\hat{A}$ ($\hat{B}$) on state $\rho$ gives rise to the probabilities of obtaining $\{a_i\}$ ($\{b_i\}$): $p_{ai}=\langle \hat{A}_i \rangle:=\Tr(\rho \hat{A}_i)$ [$p_{bi}=\langle \hat{B}_i \rangle:=\Tr(\rho \hat{B}_i)$].
We define the statistical distance, $\zeta_{\hat{A}\hat{B}}:=\sum\limits_{i}|p_{ai}-p_{bi}|$.
The statistical distance has the following properties: $\zeta_{\hat{A}\hat{B}}\geq 0$ and $\zeta_{\hat{A}\hat{B}}=0$ if two statistics coincide.

Similarly, we introduce $\{\hat{M}_i\}$ as a complete set of projective measurements on the meter state, $\{\hat{C_i}\}$ with $\hat{C_i}=\hat{U}^\dagger(\hat{I} \otimes \hat{M_i}) \hat{U}$ and $\{\hat{D_i}\}$ with $\hat{D_i}=\hat{U}^\dagger(\hat{B_i} \otimes \hat{I})\hat{U}$ as complete sets of projective measurements with respect to the noise operators $\hat{N}_{\hat{A}}$ and $\hat{N}_{\hat{B}}$, respectively. Correspondingly, one has probabilities $p_{ci}=\langle \hat{C}_i \rangle$ and $p_{di}=\langle \hat{D}_i \rangle$. In terms of statistical distance, the error and disturbance are defined by
\begin{eqnarray}
\epsilon(\hat{A})=\zeta_{\hat{A}\hat{C}}&=&\min\sum\limits_{i}|p_{a\sigma(i)}-p_{ci}|,\label{error}\\
\eta(\hat{B})=\zeta_{\hat{B}\hat{D}}&=&\min\sum\limits_{i}|p_{b\sigma(i)}-p_{di}|,\label{disturbance}
\end{eqnarray}
where the minimization takes over all index permutations $\sigma(i)$ of $i$.

We rewrite Eq. (\ref{error}) as $\epsilon(\hat{A})=\min\sum\limits_{i}|p_{a\sigma(i)}-p_{ci}|=\sum\limits_{i}|p_{ai}-p_{ci}|$, and Eq. (\ref{disturbance}) as $\eta(\hat{B})=\min\sum\limits_{i}|p_{b\sigma(i)}-p_{di}|=\sum\limits_{i}|p_{bi}-p_{di}|$ by relabelings the subindices of $P_a$ and $P_b$, respectively. Correspondingly, we have $\zeta_{\hat{A}\hat{B}}=\sum\limits_{i}|p_{ai}-p_{bi}|$ and $\zeta_{\hat{C}\hat{D}}=\sum\limits_{i}|p_{ci}-p_{di}|$.
Utilizing the triangle inequality for any metric functions, we have $\epsilon(\hat{A})+\eta(\hat{B})=\zeta_{\hat{A}\hat{C}}+\zeta_{\hat{B}\hat{D}}=\sum\limits_{i}[|p_{ai}- p_{ci}|+|p_{bi}-p_{di}|]
\geq \sum\limits_{i}[|p_{ai}-p_{bi}|-|p_{ci}-p_{di}|]= \zeta_{\hat{A}\hat{B}}-\zeta_{\hat{C}\hat{D}}$ and $\epsilon(\hat{A})+\eta(\hat{B}) \geq \sum\limits_{i}[|p_{ci}-p_{di}|-|p_{ai}-p_{bi}|]= \zeta_{\hat{C}\hat{D}}-\zeta_{\hat{A}\hat{B}}$. Therefore, we obtain $\epsilon(\hat{A})+\eta(\hat{B}) \geq |\zeta_{\hat{A}\hat{B}}-\zeta_{\hat{C}\hat{D}}|\equiv\xi_G$.
Taking into account that $\epsilon(\hat{A})=\sum\limits_{i}|p_{ai}-p_{ci}|=\sum\limits_{i}|p_{a\sigma(i)}-p_{c\sigma(i)}|$ (similarly for $\eta(\hat{B})$),
we define $\xi_{G, max}=\max \xi_G =\max |\sum\limits_{i}[|p_{a\sigma(i)}-p_{bi}|-|p_{c\sigma(i)}-p_{di}|]|$, where the maximization takes over all index permutations $\sigma(i)$.
We then reach the following theoretical result:

{\bf Theorem 1.}---The sequential measurements associated with observables $\hat{A}$ and $\hat{B}$ satisfy the following error-disturbance trade-off,
\bqa\label{thm}
\epsilon(\hat{A})+\eta(\hat{B})\geq \xi_{G, max}.
\eqa
\begin{figure}[htbp]
	\centering	
	\includegraphics[width=3.1in]{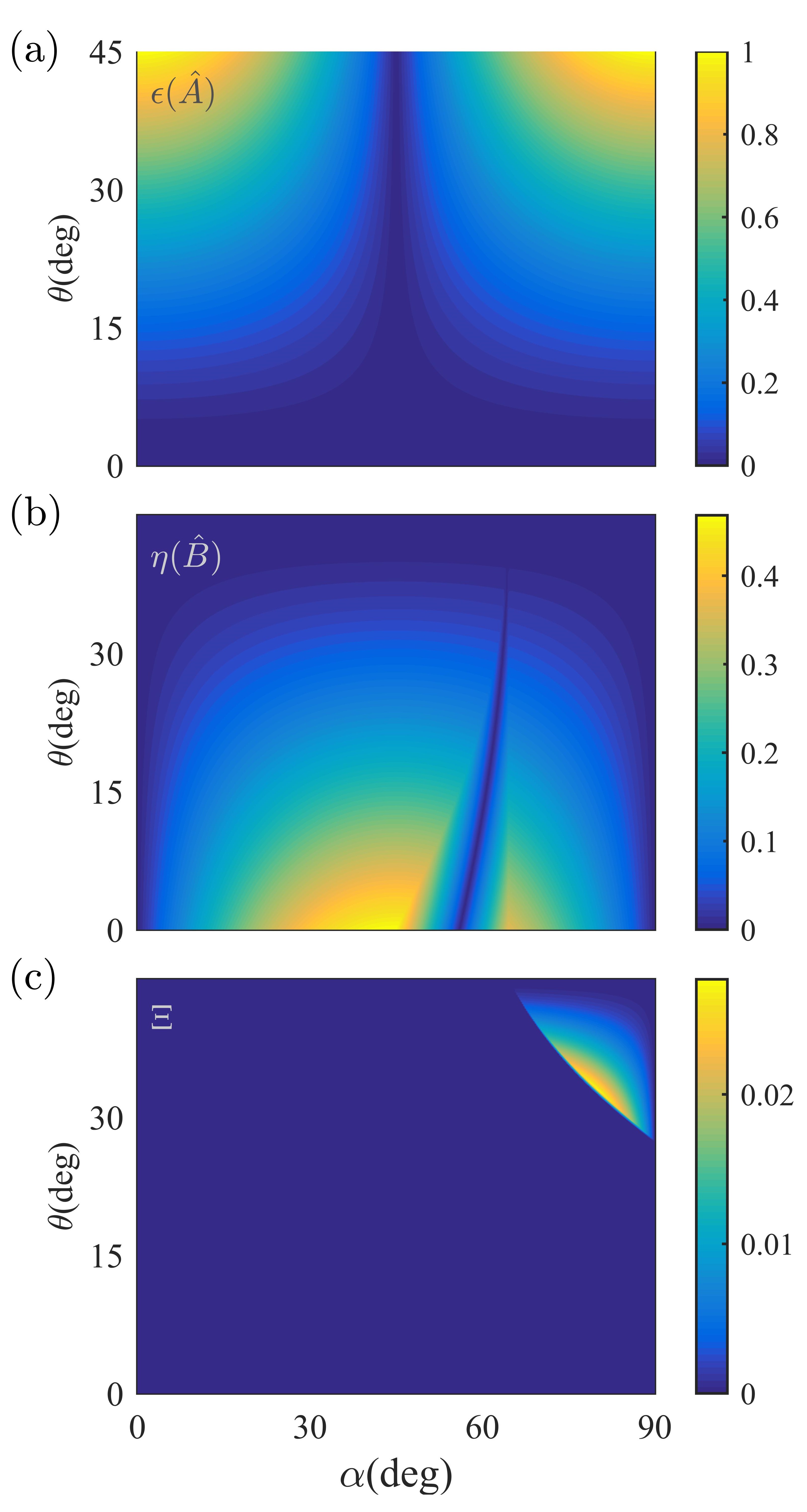}
	\caption{Computed error (a), disturbance (b), and $\Xi$ (c) for qubits under a number of experimental conditions, with amplitude indicated by the color bar. We consider here the system qubit $\ket{\Phi}_{s}=\cos{\alpha} \ket{0}_{s}+\sin{\alpha} \ket{1}_{s}$ is coupled to the meter qubit $\ket{\Phi}_{m}=\cos{\theta} \ket{0}_m+\sin{\theta} \ket{1}_m$ with a CNOT gate (see Supplementary materials). $\alpha \in[0,\pi/2]$ and $\theta\in[0,\pi/4]$. }
	\label{Fig.qubit}
\end{figure}

We note that $\hat{C}$ and $\hat{D}$ are the general measurement operators associated with $\hat{A}$ and $\hat{B}$, respectively. The lower bound $\xi_{G, max}$ may be viewed as the error induced by the general measurements. $\xi_G=0$ if $\hat{C}$ and $\hat{D}$ reduce to $\hat{A}$ and $\hat{B}$ for sharp measurements.

We notice that relation (\ref{ozawa}) is ``nonideal" \cite{PhysRevLett.112.050401,buscemipaying} at several aspects. For maximally mixed states, the right hand side (RHS) of relation (\ref{ozawa}) vanishes for any $\hat{A}$ and $\hat{B}$. It is not physically reasonable that $\epsilon(\hat{A})$ can vanish for noisy measurements, while $\eta(\hat{B})$ needs not to vanish for nondisturbing measurements. Note that in \cite{,PhysRevLett.117.140402} the right hand side of relation (\ref{ozawa}) is replaced by a stronger constant given by $D_{AB}=Tr(|\sqrt{\rho}[\hat{A},\hat{B}]\sqrt{\rho}|)/2$, which does not vanish for a maximally-mixed state. In addition, relation (\ref{ozawa}) varies with the relabelings of eigenvalues and measurement outcomes. We remark that relation (\ref{thm}) is free of these shortcomings. Relation (\ref{thm}) is also state dependent, which is different from previous state independent results from the information perspective \cite{RevModPhys.86.1261,Werner1,Werner2,Werner3,PhysRevLett.112.050401,Barchielli2018}. Furthermore, we note that our statistical distance-based measures of error and disturbance satisfy the natural and significant requirements for operational error and disturbance \cite{PhysRevA.89.052108}. We present below two theoretical analyses of relation (\ref{thm}) for qubit and qutrit, followed by an experimental verification of relation (\ref{thm}) in the photonic qubit system. We note that the Wasserstein-2 distance was used in characterizing the incompatibility of the two measurements \cite{Werner1,Werner3,PhysRevA.96.022137}, which coincides with trace distance for the qubit case up to a constant factor 2 \cite{Werner3}.

We plot the error and disturbance computed with Eqs. (2) and (3) for the measurements of a pair of noncommutative Pauli operators $\hat{A}=\hat{Z}$ and $\hat{B}=(\hat{X}+\hat{Y}+\hat{Z})/\sqrt{3}$ on qubit, respectively, in Figs. \ref{Fig.qubit}(a) and \ref{Fig.qubit}(b), which generally exhibit a complementary feature between error and disturbance (see Supplemental Material Sec. \textbf{\RNum{1}}). Defining $\Xi=\epsilon(\hat{A})+\eta(\hat{B})-\xi_{G,max}$, we plot the value of $\Xi$  in Fig. \ref{Fig.qubit}(c). It is evident that $\Xi\geq0$, hence confirming relation (\ref{thm}). We note that if $\hat{A}$ and $\hat{B}$ in their Bloch representations $\hat{A}=\vec{a}\cdot\vec{\sigma}$ and $\hat{B}=\vec{b}\cdot\vec{\sigma}$ are orthogonal, $\vec{a}\cdot\vec{b}=0$, we have $\Xi=0$, the complementary property is more pronounced (see Supplemental Material Sec. \textbf{\RNum{1}}). The stripe appearing on the lower part of Fig. \ref{Fig.qubit}(b) is due to the nonorthogonal part, which may be interpreted as the partial information gained in the measurement of $\hat{B}$ that is compatible with the measurement of $\hat{A}$.

We plot the computed error and disturbance for the measurements of a pair of noncommutative angular momentum observables on qutrit, with $\hat{A}=\hat{L}_z$ and $\hat{B}=\hat{L}_x$, $[\hat{L}_z, \hat{L}_x]=i\hat{L}_y$, respectively in Figs. \ref{Fig.qutrit}(a) and \ref{Fig.qutrit}(b), which exhibits a clear complementary feature (see Supplemental Material Sec. \textbf{\RNum{1}}). We plot the value of $\Xi$  in Fig. \ref{Fig.qutrit}(c). It is evident that $\Xi\geq0$, hence confirming relation (\ref{thm}).

\begin{figure}[htbp]
	\centering	
	\includegraphics[width=3.1in]{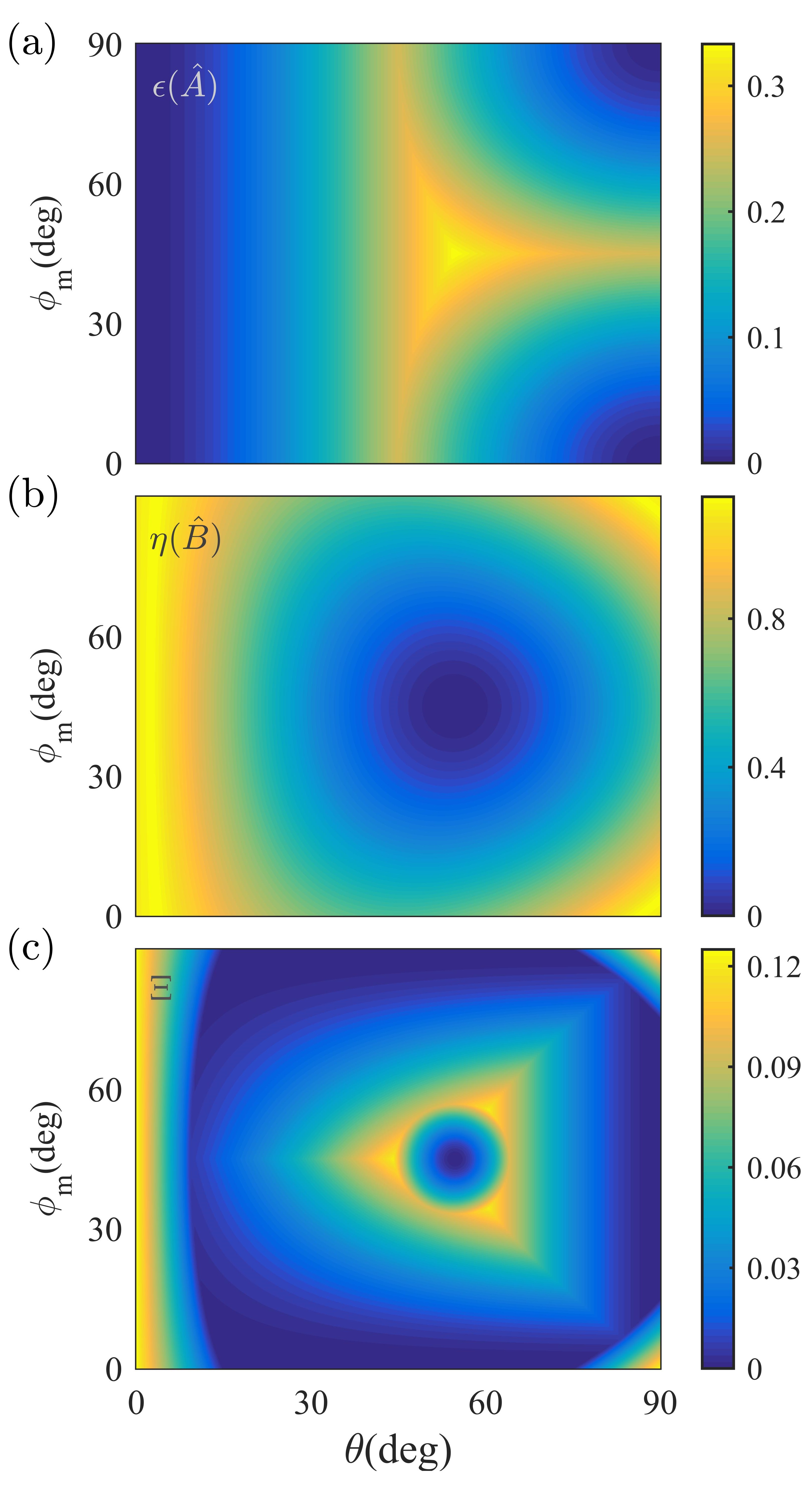}
	\caption{Computed error (a), disturbance (b), and $\Xi$ (c) for qutrits under a number of experimental conditions, with amplitude indicated by the color bar. We consider here the system qutrit $\ket{\Phi}_{s}=\sin{\alpha} \cos{\phi_{s}} \ket{0}_{s}+\sin{\alpha} \sin{\phi_{s}} \ket{1}_{s}+\cos{\alpha} \ket{2}_{s}$ is coupled to the meter qutrit $\ket{\Phi}_{m}=\sin{\theta} \cos{\phi_{m}} \ket{0}_m+\sin{\theta} \sin{\phi_{m}} \ket{1}_m+\cos{\theta} \ket{2}_{m}$ with a generalized control gate $\hat{U}$~\cite{qutrits} (see Supplementary materials). $\alpha=\phi_s=\pi/4$, $\theta \in[0,\pi/2]$ and $\phi_m\in[0,\pi/2]$. }
	\label{Fig.qutrit}
\end{figure}

\begin{figure}[htbp]
	\centering
	\includegraphics[width=3.4in]{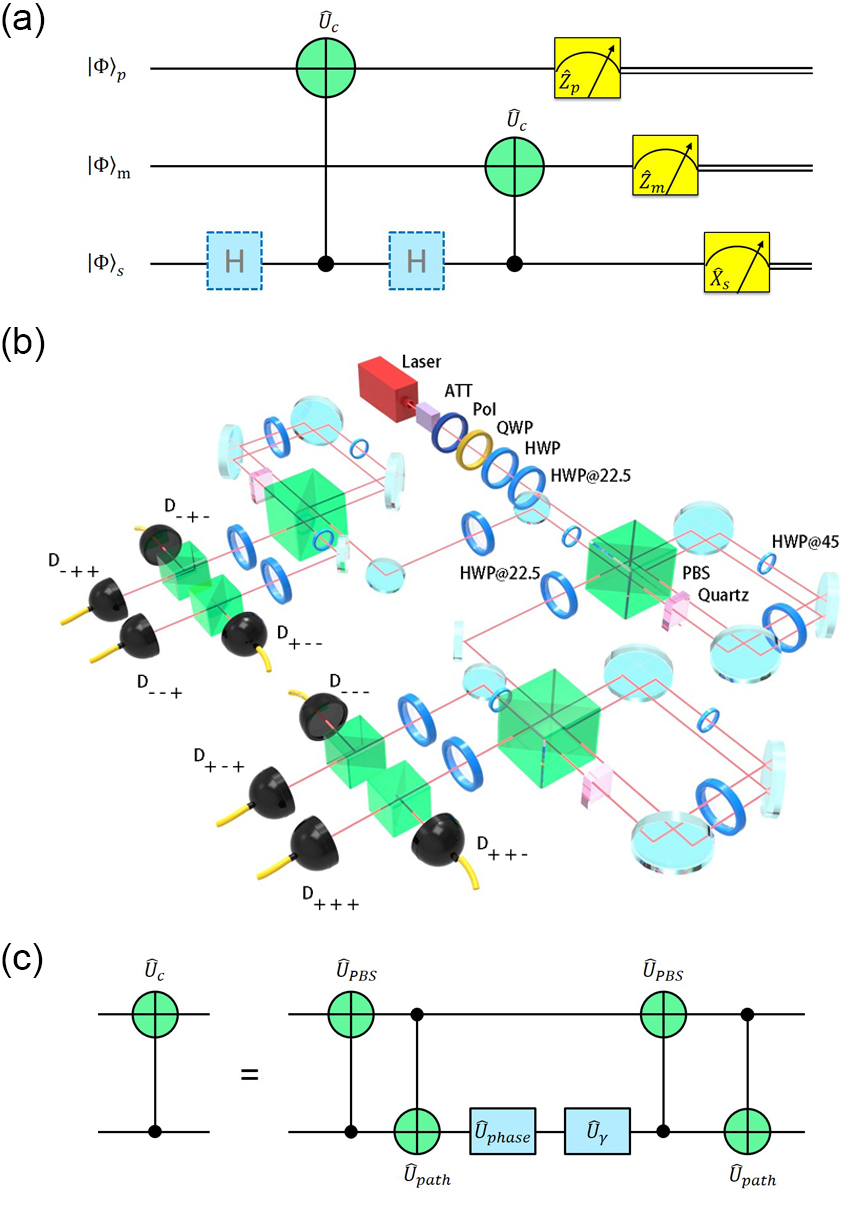}
	\caption{(a) Quantum circuit model of measuring observable $\hat{X}$ ($\hat{Z}$) with (without) applying Hadamard gates $\hat{H}$. The system qubit is sequentially coupled to the probe qubit and the meter qubit by CNOT gates, with measurement outcomes $z_p$, $z_m$ and $x_s$, respectively. (b) We prepare an arbitrary system qubit on the Bloch sphere by passing the attenuated laser beam through a polarizer and a pair of half (HWP) and quarter (QWP) wave plates. A HWP is set to $0^\circ$ oriented from the vertical for the measurement of observable $\hat{Z}$ or to $22.5^\circ$ (HWP@22.5) to implement a Hadamard gate $\hat{H}$ for the measurement of observable $\hat{X}$. We first couple the system qubit to probe qubit by a Sagnac interferometer. This coupling yields two identical outputs in two paths, and each couples to a meter qubit via a Sagnac interferometer. This results in four outputs, which create eight outputs after polarization analysis. Each of the eight outputs is detected by a single photon detector $D$. (c) A quantum circuit illustration of implementing the coupling ($\hat{U}_C$) with a Sagnac interferometer, $\hat{U}_{C}=\hat{U}_{path} \hat{U}_{PBS} \hat{U}_{\gamma} \hat{U}_{phase} \hat{U}_{path} \hat{U}_{PBS}$.}
	\label{fig:setup}
\end{figure}

\begin{figure}[htbp]
	\centering
	\includegraphics[width=3.2in]{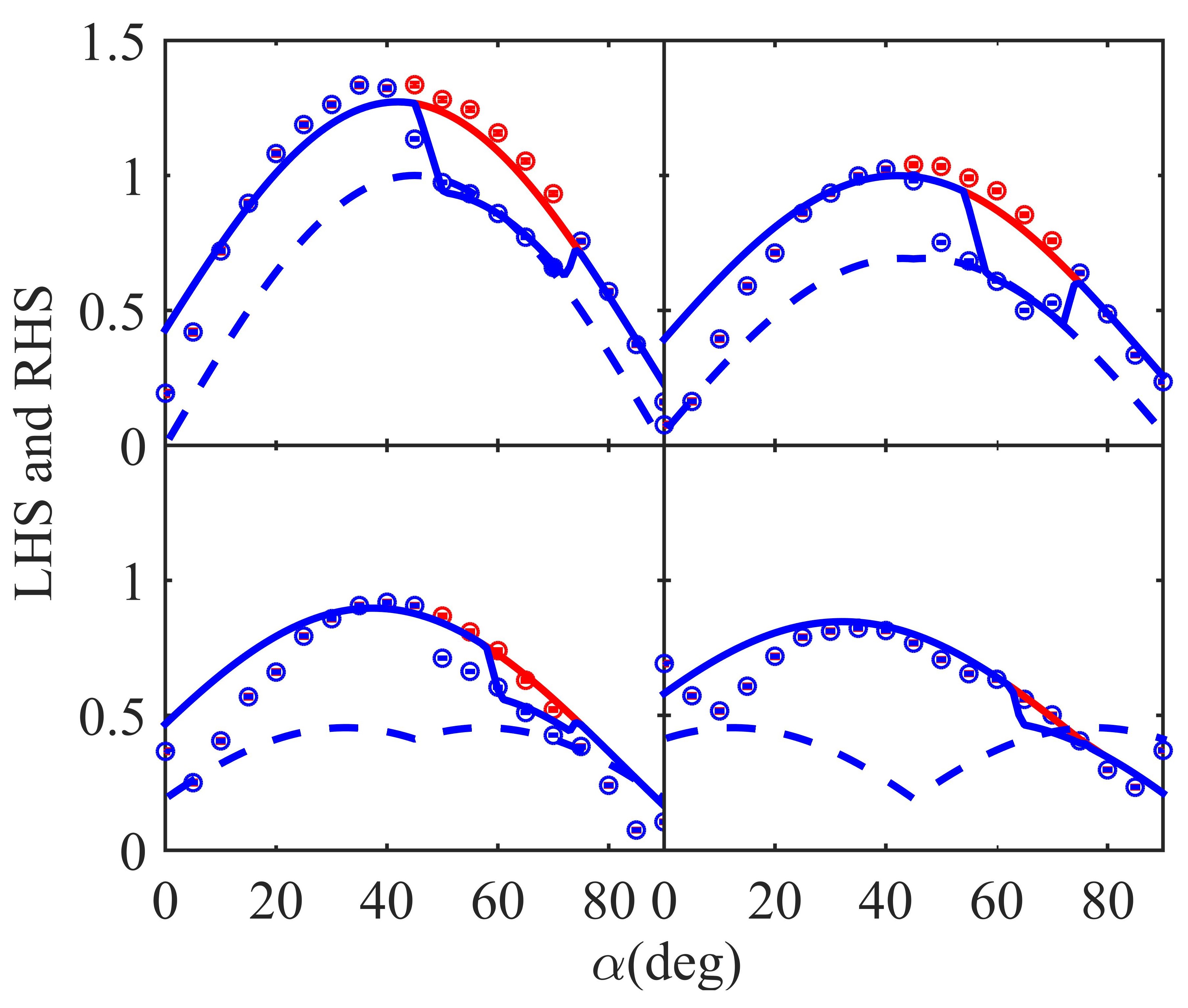}
	\caption{Experimental measurements of the LHS and RHS of relation (\ref{thm}). The single photon system qubit is given by $\ket{\Phi}_{s}=\cos\alpha\ket{0}_{pol}+\sin\alpha\ket{1}_{pol}$. From (a) to (d), the angle $\theta$ in the measurement strength $\cos2\theta$ is set to $0^\circ$, $9^\circ$, $18^\circ$, and $27^\circ$, respectively. LHS: red, RHS: blue. Dashed line: ideal theory, smooth line: theory corrected with the imperfection in PBS, circles: experimental results. The error bars stand for one standard deviation, assuming Poissonian statistics.}
	\label{fig:HV}
\end{figure}

We now present an experimental verification of relation (\ref{thm}) with the measurements of a pair of noncommutative Pauli operators $\hat{A}=\hat{Z}$ and $\hat{B}=\hat{X}$ on a photonic qubit system. The quantum circuit for the experimental implementation is given in Fig. \ref{fig:setup}(a) \cite{Lund}. The system qubit $\ket{\Phi}_s=\cos\alpha\ket{0}_s+e^{i\phi}\sin\alpha\ket{1}_s$ is sequentially coupled (via a unitary operation $\hat{U}_C$) to a probe qubit, $\ket{\Phi}_p=\gamma\ket{0}_p+\bar{\gamma}\ket{1}_p$, and a meter qubit, $\ket{\Phi}_m=\cos\theta\ket{0}_m+\sin\theta\ket{1}_m$. All three states are normalized. The quantum circuit yields eight joint measurement outcomes $\in z_p \otimes z_m\otimes x_s$, with $z_p$, $z_m$ and $x_s$ $=\{+,-\}$, which are completely described by a set of positive-operator-valued measures (POVMs) $\{\hat{\Pi}_{jkl}\}$, with $j$, $k$ and $l\in \{+,-\}$.
Denote $\hat{\Pi}_j=\Sigma_{kl}\hat{\Pi}_{jkl}$,  $\hat{\Pi}_k=\Sigma_{jl}\hat{\Pi}_{jkl}$ and $\hat{\Pi}_l=\Sigma_{jk}\hat{\Pi}_{jkl}$, with $\Sigma_j\hat{\Pi}_j=\Sigma_k\hat{\Pi}_k=\Sigma_l\hat{\Pi}_l=1$.
Here the one-to-one correspondence is given between $z_p$ and $\hat{\Pi}_j$, $z_m$ and $\hat{\Pi}_k$, and $x_s$ and $\hat{\Pi}_l$. Associated with the POVMs, the probabilities $\{p_{a\pm}\}$, $\{p_{b\pm}\}$, $\{p_{c\pm}\}$, and $\{p_{d\pm}\}$ in relation (\ref{thm}) are given by (see Supplemental Material Sec. \textbf{\RNum{2}}),
\bqa
\begin{aligned}
p_{a\pm}&=\frac{1}{2}(1\pm\frac{\langle\hat{\Pi}_{j=+}^a\rangle-\langle\hat{\Pi}_{j=-}^a\rangle}{2\gamma^2-1}),&\\ p_{b\pm}&=\frac{1}{2}(1\pm\frac{\langle\hat{\Pi}_{j=+}^b\rangle-\langle\hat{\Pi}_{j=-}^b\rangle}{2\gamma^2-1}),&\\ p_{c\pm}&=\langle\hat{\Pi}_{k=\pm}\rangle=\frac{1}{2}(1\pm\cos2\theta \langle\hat{Z}\rangle),&\\
p_{d\pm}&=\langle\hat{\Pi}_{l=\pm}\rangle=\frac{1}{2}(1\pm\sin2\theta \langle\hat{X}\rangle),&
\end{aligned}
\eqa
where $\hat{\Pi}_{j=\pm}^a=\frac{1}{2}[\hat{I}\pm(2\gamma^2-1)\hat{Z}]$ and
$\hat{\Pi}_{j=\pm}^b=\frac{1}{2}[\hat{I}\pm(2\gamma^2-1)\hat{X}]$. It is straightforward to show that the error and disturbance trade-off relation (\ref{thm}) holds tight (dashed lines in Fig. \ref{fig:HV}).

{\it Experimental implementation}.---As shown in Fig. \ref{fig:setup}(b), we encode the system, probe and meter qubits using the polarization and path degrees of freedom of single photons, respectively. We first attenuate the emission of a continuous wave, distributed feedback laser with a linewidth of 2 MHz at 1560 nm to approximate a single photon source, $|\beta\rangle\approx|0\rangle+\beta|1\rangle$ with $|\beta|^2 \ll1$. We then pass the single photons through a polarizer (with polarization extinction better than $10^5:1$). With a pair of half and quarter wave plates, we can create arbitrarily a polarization qubit on the Bloch sphere as the system qubit, $\ket{\Phi}_{s}=\cos\alpha\ket{0}_{pol}+e^{i\phi}\sin\alpha\ket{1}_{pol}$, where $\alpha$ is the angle of the fast axis of a HWP oriented from the vertical, states $\ket{0}_{pol}$ and $\ket{1}_{pol}$ stand for horizontal and vertical polarization states $\ket{H}$ and $\ket{V}$, respectively, and $\phi$ is the phase. The probe qubit, $|\Phi\rangle_p=\gamma|0\rangle_{path}+\bar{\gamma}|1\rangle_{path}$, is encoded with the path degree of freedom of single photons, with $|0\rangle_{path}$ for clockwise and $|1\rangle_{path}$ for counterclockwise propagation states in the Sagnac interferometer.

We implement the unitary coupling $\hat{U}_{C}$ of the system qubit, $\ket{\Psi}_{s}=(\cos\alpha\ket{0}_{pol}+e^{i\phi}\sin\alpha\ket{1}_{pol})|0\rangle_{path}$, to the probe qubit with a Sagnac interferometer
\cite{Lund,PhysRevLett.114.070402}, with $\hat{U}_{C}=\hat{U}_{path} \hat{U}_{PBS} \hat{U}_{\gamma} \hat{U}_{phase} \hat{U}_{path} \hat{U}_{PBS}$ illustrated in Fig. \ref{fig:setup}(c). Both $\hat{U}_{PBS}$ and $\hat{U}_{path}$ are CNOT gates. While $\hat{U}_{PBS}$ is implemented with a polarizing beam splitter (PBS) with polarization qubit as the control and the path qubit as the target, $\hat{U}_{path}$ is implemented with a half wave plate oriented at $45^o$ (HWP@45) from the vertical with the path qubit as the control and the polarization qubit as the target. $\hat{U}_{phase}$ is a $\hat{Z}$-phase gate, where we simply use the fact that single photons in state $|V\rangle$ acquire a $\pi-$ phase with respect to photons in state $|H\rangle$ upon reflection on a mirror. $\hat{U}_{\gamma}$ describes the coupling of the polarization qubit to path qubit by a HWP.

We subsequently couple the system qubit to the meter qubit with two Sagnac interferometers to account for the two paths (see Supplemental Material Sec. \textbf{\RNum{3}}). Note that a quartz plate is used to null the phase difference between the clockwise and counterclockwise propagation states in the interferometer, and HWPs are inserted to realize the Hadamard gate for the measurement of $\hat{X}$.

We use InGaAs single photon detectors [labeled by $D_{jkl}$ in Fig. \ref{fig:setup}(b), with $j,k,l=\pm$] to detect single photons from the eight output ports in the experiment, with the gating window of the detectors set to 2 ns and the duty cycle set to 1 $\mu$s to reduce background noise. With the number of single photon detection events at the eight output ports, $N_{jkl}$, we compute the probabilities as $P_j=\frac{\sum_{kl}N_{jkl}}{N}$, $P_k=\frac{\sum_{jl}N_{jkl}}{N}$, and $P_l=\frac{\sum_{jk}N_{jkl}}{N}$, respectively, where $N=\sum_{jkl}N_{jkl}$. We then obtain
$p_{a\pm}( p_{b\pm})=\frac{1}{2}(1\pm\frac{P_{j=+}-P_{j=-}}{2\gamma^2-1})$ for the measurement of $\hat{Z}$ ($\hat{X}$),
$p_{c\pm}=\langle\hat{\Pi}_{k=\pm}\rangle=P_{k=\pm}$ and
$p_{d\pm}=\langle\hat{\Pi}_{l=\pm}\rangle=P_{l=\pm}$.

Note that $\gamma$ varies from $\sqrt{2}/2$ for no coupling to $\gamma =1$ for projective (sharp) measurement. We set $\gamma=0.766$ to work in the weak measurement limit in the experiment.

For the linearly polarized system qubit, $\ket{\Phi}_{s}=\cos\alpha\ket{0}_{pol}+\sin\alpha\ket{1}_{pol}$, we vary in the experiment the linear polarization of the system qubit from state $|H\rangle$ to state $|V\rangle$ and the strength to couple the system qubit to the meter qubit from zero for no coupling to one for projective measurement. As we noted earlier in the text, we have $\Xi=0$ in this case. For a better illustration, we plot the values of the LHS and RHS of relation (\ref{thm}) (red and blue circles), respectively,  in Fig. \ref{fig:HV}. They generally coincide with each other with a few exceptions that the LHS is greater than RHS. By incorporating the imperfection of PBS, i.e., the imperfect PBS reflects (transmits) a very small percentage of single photons in state $|H\rangle$ ($|V\rangle$), we theoretically reproduce the experimental observations (smooth line), hence verifying the error-disturbance relation (\ref{thm}). (see Supplemental Material Sec. \textbf{\RNum{4}} for details and \textbf{\RNum{5}} for results on the circularly polarized system qubit).

{\it Conclusion}.---We theoretically derived and experimentally verified that the summation of error and disturbance quantified by the statistical distance is lower bounded by a tight inequality. This new trade-off relation is free of the shortcomings of relation (\ref{ozawa}) and the lower bound is state dependent. We anticipate that our work may stimulate further investigations on quantum uncertainties and applications in measurement science.

This work has been supported by the National Key R\&D Program of China (2017YFA0303900, 2017YFA0304000), National Fundamental Research Program, Chinese Academy of Sciences, National Natural Science Foundation of China (11675113) and Key Project of Beijing Municipal Commission of Education (KZ201810028042).\\

Y-L. M., and Z-H. M. contributed equally to this work.\\

\onecolumngrid

\subsection*{\textbf{\large Supplemental Materials: Error-Disturbance trade-off in Sequential Quantum Measurements}}

	\section{The theoretical results of $\Xi$}
	According to the main text, we consider a given quantum state $\rho_s$, meter state $\rho_m$, and two observables, $\hat{A}=\sum\limits_{i}a_i\hat{A}_{i}$ and $\hat{B}=\sum\limits_{i}b_i\hat{B}_{i}$, where $\{\hat{A}_i\}$ and $\{\hat{B}_i\}$ ($\{a_i\}$ and $\{b_i\}$) are the projective measurements with respect to the eigenvectors (eigenvalues) of $\hat{A}$ and $\hat{B}$, respectively. The measurement $\hat{A}$ ($\hat{B}$) on state $\rho$ gives rise to the probabilities of obtaining $\{a_i\}$ ($\{b_i\}$): $p_{ai}=\langle \hat{A}_i \rangle:=\Tr(\rho \hat{A}_i)$ ($p_{bi}=\langle \hat{B}_i \rangle:=\Tr(\rho \hat{B}_i)$).
	
	Following the indirect measurement model \cite{Lund}, $\{\hat{C_i}\}$ with $\hat{C_i}=\hat{U}^\dagger(\hat{I} \otimes \hat{M_i})\hat{U}$ and $\{\hat{D_i}\}$ with $\hat{D_i}=\hat{U}^\dagger(\hat{B_i} \otimes \hat{I})\hat{U}$ are complete sets of general measurements with respect to sharp measurements $\hat{A}$ and $\hat{B}$, respectively. Here  we set  $\hat{M}=\hat{A}$ with $\{\hat{M}_i\}$ as a complete set of projective measurements on the meter state. In the following we examine our error-disturbance trade-off for qubits and qutrits.
	
	\subsection{The theoretical results of $\Xi$ for qubits}
	We consider the system state $\ket{\Phi}_s=\cos\alpha\ket{0}_s+e^{i\phi}\sin\alpha\ket{1}_s$, so the density matrix is $\rho_s=({1+\vec{r}_s\cdot\vec{\sigma}})/{2}$, where $\vec{r}_s=(\sin{2\alpha}\cos\phi,\sin{2\alpha}\sin\phi,\cos2\alpha)$ and $\vec{\sigma}=(\hat{X},\hat{Y},\hat{Z})$ are the three Pauli matrices. The meter state is $\ket{\Phi}_{m}=\cos\theta\ket{0}_{m}+\sin\theta\ket{1}_{m}$, with the density matrix $\rho_m=({1+\vec{r}_m\cdot\vec{\sigma}})/{2}$ and $\vec{r}_m=(\sin{2\theta},0,\cos2\theta)$. The unitary operater $\hat{U}$ is a Controlled-NOT (CNOT) gate, $\hat{U}=\ket{0}_{s}{}_{s}\bra{0}\otimes \hat{I}_m+\ket{1}_{s}{}_{s}\bra{1}\otimes\hat{X}_m$. We choose two observables $\hat{A}=\vec{a}\cdot\vec{\sigma}$ and $\hat{B}=\vec{b}\cdot\vec{\sigma}$ with $\vec{a}=(0,0,1)$ and $\vec{b}=(\sin\theta_b\cos\varphi,\sin\theta_b\sin\varphi,\cos\theta_b)$. Consequently, we have 
	\beq
	\begin{aligned}
		p_{a\pm}&=\Tr(\rho_{s}\hat{A}_{\pm})=\frac{1}{2}(1\pm\langle\hat{A}\rangle),&\\
		p_{b\pm}&=\Tr(\rho_{s}\hat{B}_{\pm})=\frac{1}{2}(1\pm\langle\hat{B}\rangle),&\\
		p_{c\pm}&=\Tr((\rho_{s}\otimes\rho_{m})\hat{C}_{\pm})=\frac{1}{2}(1\pm\cos2\theta \langle\hat{A}\rangle),&\\
		p_{d\pm}&=\Tr((\rho_{s}\otimes\rho_{m})\hat{D}\pm)=\frac{1}{2}\{1\pm(\sin2\theta\langle\hat{B}\rangle+(1-\sin2\theta)(\vec{b}\cdot\vec{a})\langle\hat{A}\rangle)\},&
	\end{aligned}
	\eeq
	where $\langle\hat{A}\rangle=\vec{a}\cdot\vec{r_s}$ and $\langle\hat{B}\rangle=\vec{b}\cdot\vec{r_s}$. Correspondingly, the measurement errors $\epsilon(\hat{A})$ ($\eta(\hat{B})$) of the observables $\hat{A}$ ($\hat{B}$) are given by
	\beq
	\begin{aligned}
		\epsilon(\hat{A})&=\zeta_{\hat{A}\hat{C}} =\min\limits\sum_{i}\limits|p_{ai}-p_{ci}|=(1-\cos2\theta)|\langle\hat{A}\rangle|,&\\
		\eta(\hat{B})&=\zeta_{\hat{B}\hat{D}}=\min\limits\sum\limits_{i}|p_{bi}-p_{di}|=\min\{\eta_1(\hat{B}), \eta_2(\hat{B})\},&\\
	\end{aligned}
	\eeq
	where
	\beq
	\begin{aligned}	
		\eta_1(\hat{B})&=|p_{b+}-p_{d+}|+|p_{b-}-p_{d-}|=(1-\sin2\theta)|\langle\hat{B}\rangle-(\vec{b}\cdot\vec{a})\langle\hat{A}\rangle|,&\\
		\eta_2(\hat{B})&=|p_{b+}-p_{d-}|+|p_{b-}-p_{d+}|=|(1+\sin2\theta)\langle\hat{B}\rangle-(1-\sin2\theta)(\vec{b}\cdot\vec{a})\langle\hat{A}\rangle|.&\\
	\end{aligned}
	\eeq
	\begin{figure}[htbp]
		\centering
		\includegraphics[width=8cm]{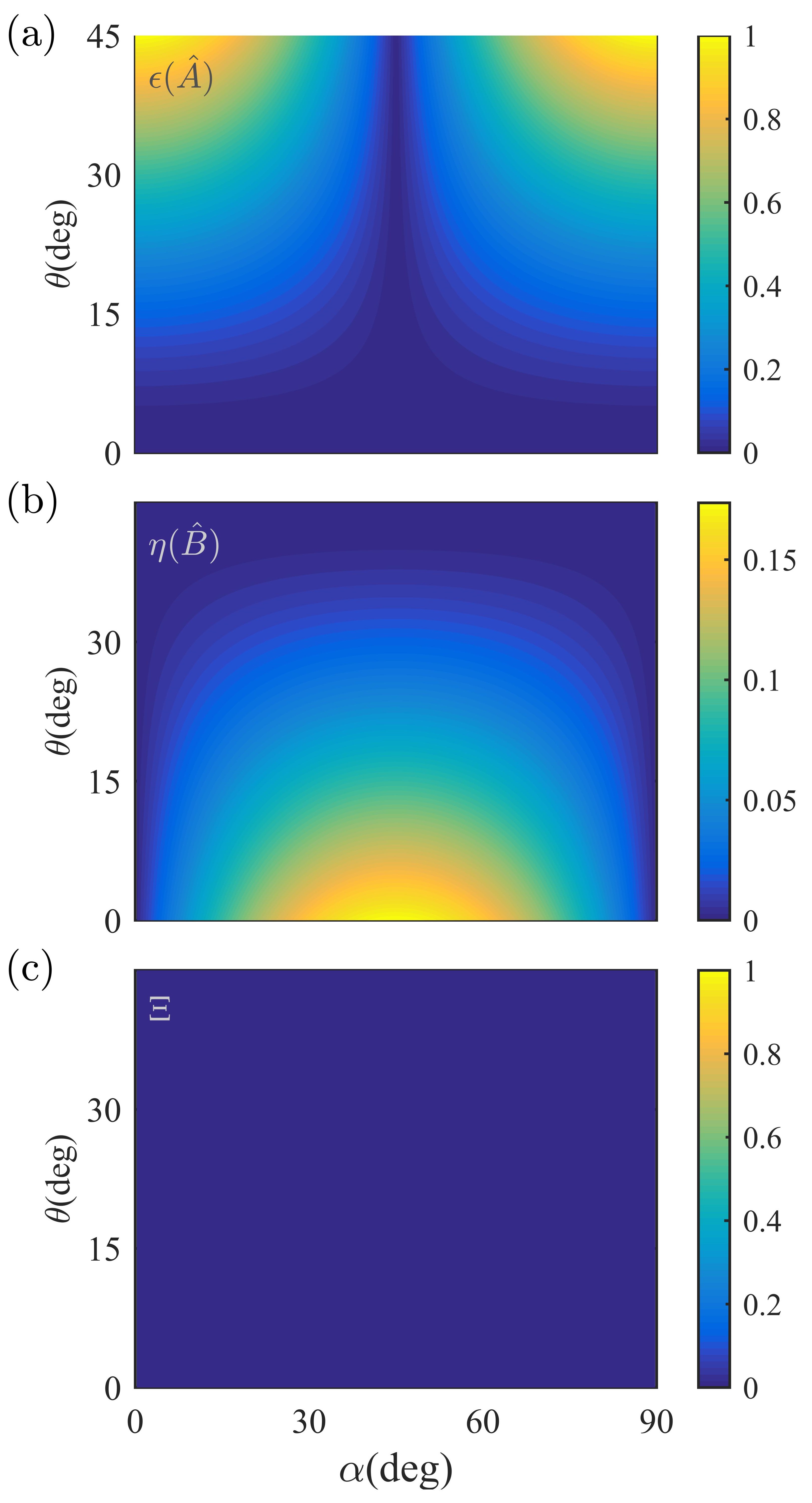}
		\caption{Computed error (a), disturbance (b), and $\Xi$ (c) for qubits under a number of experimental conditions, with amplitude indicated by the color bar. We consider the system qubit $\ket{\Phi}_{s}=\cos{\alpha} \ket{0}_{s}+\sin{\alpha} \ket{1}_{s}$ is coupled to the meter qubit $\ket{\Phi}_{m}=\cos{\theta} \ket{0}_m+\sin{\theta} \ket{1}_m$ with a CNOT gate. We set $\hat{A}=\hat{Z}$ and $\hat{B}=\hat{X}$. $\alpha \in[0,\pi/2]$ and $\theta\in[0,\pi/4]$. }
		\label{fig:FIGS1}
	\end{figure}
	
	For $\eta(\hat{B})=\eta_1(\hat{B})$, we have 
	\beq
	\begin{aligned}
		\xi_{G, max}&=\max \xi_G =\max\{\xi_{G1},\xi_{G2}\},\\
		\xi_{G1} &=||p_{a+}-p_{b+}|+|p_{a-}-p_{b-}|-|p_{c+}-p_{d+}|-|p_{c-}-p_{d-}||,&\\
		&=||\langle\hat{A}\rangle-\langle\hat{B}\rangle|-|\cos2\theta\langle\hat{A}\rangle-\sin2\theta\langle\hat{B}\rangle-(1-\sin2\theta)(\vec{b}\cdot\vec{a})\langle\hat{A}\rangle||,&\\
		\xi_{G2} &=||p_{a+}-p_{b-}|+|p_{a-}-p_{b+}|-|p_{c+}-p_{d-}|-|p_{c-}-p_{d+}||,&\\
		&=||\langle\hat{A}\rangle+\langle\hat{B}\rangle|-|\cos2\theta\langle\hat{A}\rangle+\sin2\theta\langle\hat{B}\rangle+(1-\sin2\theta)(\vec{b}\cdot\vec{a})\langle\hat{A}\rangle||.\\
	\end{aligned}
	\eeq
	We then have 
	\bqa
	\begin{aligned}
		\Xi=\epsilon(\hat{A})+\eta(\hat{B})-\xi_{G, max}=\min\{(\epsilon(\hat{A})+\eta_1(\hat{B})-\xi_{G,1}),(\epsilon(\hat{A})+\eta_1(\hat{B})-\xi_{G,2})\}=\min\{\Xi_1,\Xi_2\}.\\
	\end{aligned}
	\eqa
	
	For $\eta(\hat{B})=\eta_2(\hat{B})$, we have
	\beq
	\begin{aligned}
		\xi_{G, max}&=\max \xi_G =\max\{\xi_{G3},\xi_{G4}\},\\
		\xi_{G3} &=||p_{a+}-p_{b+}|+|p_{a-}-p_{b-}|-|p_{c+}-p_{d-}|-|p_{c-}-p_{dc+}||,&\\
		&=||\langle\hat{A}\rangle-\langle\hat{B}\rangle|-|\cos2\theta\langle\hat{A}\rangle-\sin2\theta\langle\hat{B}\rangle-(1-\sin2\theta)(\vec{b}\cdot\vec{a})\langle\hat{A}\rangle||,&\\
		\xi_{G4} &=||p_{a+}-p_{b-}|+|p_{a-}-p_{b+}|-|p_{c+}-p_{d+}|-|p_{c-}-p_{d-}||,&\\
		&=||\langle\hat{A}\rangle+\langle\hat{B}\rangle|-|\cos2\theta\langle\hat{A}\rangle+\sin2\theta\langle\hat{B}\rangle+(1-\sin2\theta)(\vec{b}\cdot\vec{a})\langle\hat{A}\rangle||,\\
	\end{aligned}
	\eeq
	we then have
	\bqa
	\begin{aligned}
		\Xi=\epsilon(\hat{A})+\eta(\hat{B})-\xi_{G, max}=\min\{(\epsilon(\hat{A})+\eta_2(\hat{B})-\xi_{G,3}),(\epsilon(\hat{A})+\eta_2(\hat{B})-\xi_{G,4})\}=\min\{\Xi_3,\Xi_4\}.\\ 
	\end{aligned}
	\eqa
	We now consider two cases:
	
	$(1)$ $\vec{b}\cdot\vec{a}=0$. $\Xi$ reduces to $\min\{\Xi_1,\Xi_2\}$, and we have
	\bqa
	\left\{
	\begin{aligned}
		& \Xi_1\geq 0 \quad and \quad  \Xi_2=0, & \langle\hat{A}\rangle\geq 0 \quad and \quad \langle\hat{B}\rangle\geq 0,\\
		& \Xi_1\geq 0 \quad and \quad  \Xi_2=0, & \langle\hat{A}\rangle\le 0 \quad and \quad \langle\hat{B}\rangle\le 0,\\
		& \Xi_2\geq 0 \quad and \quad \Xi_1=0, & \langle\hat{A}\rangle\geq 0 \quad and \quad \langle\hat{B}\rangle\le 0,\\
		& \Xi_2\geq 0 \quad and \quad \Xi_1=0, & \langle\hat{A}\rangle\le 0 \quad and \quad \langle\hat{B}\rangle\geq 0.\\
	\end{aligned}
	\right.
	\eqa
	It is obvious that $\Xi=0$. Hence the relation (4) is tight .
	
	We plot the error and disturbance computed for the measurements of a pair of noncommutative Pauli operators $\hat{A}=\hat{Z}$ and $\hat{B}=\hat{X}$ on qubit respectively in Fig. \ref{fig:FIGS1}(a) and (b), which generally exhibits a complementary feature between error and disturbance. We plot the value of $\Xi$  in Fig. \ref{fig:FIGS1}(c). It is evident that $\Xi=0$, hence confirming our relation.   
	
	$(2)$ $\vec{b}\cdot\vec{a}\neq 0$. We set $\hat{A}=\hat{Z}$ and $\hat{B}=(\hat{X}+\hat{Y}+\hat{Z})/{\sqrt{3}}$, see the main text.
	
	\subsection{The theoretical results of $\Xi$ for qutrits}
	We consider the sets of angular momentum operators \{$\hat{L}_x$, $\hat{L}_y$, $\hat{L}_z$\}, with $[\hat{L}_i, \hat{L}_j]=\epsilon_{ijk}\hat{L}_k$. We set $\hat{A}=\hat{L}_z$ with eigenvalues 1, 0 and -1. The corresponding eigenvectors
	\beq
	\begin{aligned}
		\centering
		\hat{a}_1=\left(\begin{array}{c} 1 \\ 0 \\ 0  \end{array}\right),~~~
		\hat{a}_2=\left(\begin{array}{c} 0 \\ 1 \\ 0  \end{array}\right),~~~
		\hat{a}_3=\left(\begin{array}{c} 0 \\ 0 \\ 1  \end{array}\right),
		\centering
	\end{aligned}
	\eeq
	we then have 
	\beq
	\begin{aligned}
		\centering
		\hat{L}_x=\left(\begin{array}{ccc} 0 & \frac{1}{\sqrt{2}}& 0 \\ \frac{1}{\sqrt{2}}& 0 & \frac{1}{\sqrt{2}} \\ 0 & \frac{1}{\sqrt{2}}& 0 \\ \end{array}\right),
		\hat{L}_y=\left(\begin{array}{ccc} 0 & -\frac{1}{\sqrt{2}}i& 0 \\ \frac{1}{\sqrt{2}}i& 0 & -\frac{1}{\sqrt{2}}i \\ 0 & \frac{1}{\sqrt{2}}i& 0 \\ \end{array}\right),
		\hat{L}_z=\left(\begin{array}{ccc} 1 & 0 & 0 \\ 0 & 1 & 0\\ 0 & 0 & -1 \\ \end{array}\right).	
		\centering
	\end{aligned}
	\eeq
	We consider $\hat{B}=\hat{L}_x$, with eigenvalues 1, 0 and -1, and eigenvectors
	\beq
	\begin{aligned}
		\centering
		\hat{b}_1=\left(\begin{array}{c} \frac{1}{2} \\[2mm] \frac{1}{\sqrt{2}}\\[2mm] \frac{1}{2} \end{array}\right),
		\hat{b}_2=\left(\begin{array}{c} -\frac{1}{\sqrt{2}}\\[2mm]0\\[2mm] \frac{1}{\sqrt{2}} \end{array}\right),
		\hat{b}_3=\left(\begin{array}{c} \frac{1}{2} \\[2mm] -\frac{1}{\sqrt{2}}\\[2mm] \frac{1}{2} \end{array}\right),
		\centering
	\end{aligned}
	\eeq
	Accordingly, we take the measurements on the meter state $\hat{M}=\hat{A}$ and the unitary operater $\hat{U}$ \cite{qutrits} as:	
	\beq
	\begin{aligned}
		\centering
		\hat{U}=\left(\begin{array}{ccccccccc} 1 & 0 & 0 & 0 & 0 & 0 & 0 & 0 & 0\\ 0 & 1 & 0 & 0 & 0 & 0 & 0 & 0 & 0\\ 0 & 0 & 1 & 0 & 0 & 0 & 0 & 0 & 0\\ 0 & 0 & 0 & 0 & 0 & 1 & 0 & 0 & 0\\ 0 & 0 & 0 & 1 & 0 & 0 & 0 & 0 & 0\\ 0 & 0 & 0 & 0 & 1 & 0 & 0 & 0 & 0\\ 0 & 0 & 0 & 0 & 0 & 0 & 0 & 1 & 0\\ 0 & 0 & 0 & 0 & 0 & 0 & 0 & 0 & 1\\ 0 & 0 & 0 & 0 & 0 & 0 & 1 & 0 & 0 \end{array}\right).
		\centering
	\end{aligned}
	\eeq
	
	We consider a system state $\ket{\Phi}_{s}=f_1
	\ket{0}_{s}+f_2\ket{1}_{s}+f_3\ket{2}_{s}=\sin{\alpha}\cos{\phi_{s}}\ket{0}_{s}+e^{i\chi_{12}}\sin{\alpha}\sin{\phi_{s}}\ket{1}_{s}+e^{i\chi_{13}}\cos{\alpha}\ket{2}_{s}$, where $0\leq\alpha,\phi_{s}\leq\pi/2$, $0\leq\chi_{12},\chi_{13}\leq 2\pi$, and ${\left| f_1 \right|}^2+{\left|f_2 \right|}^2+{\left|f_3 \right|}^2=1$. The meter state is $\ket{\Phi}_{m}=\sin{\theta}\cos{\phi_{m}}\ket{0}_{m}+\sin{\theta}\sin{\phi_{m}}\ket{1}_{m}+\cos{\theta}\ket{2}_{m}$,  where $0\leq\theta,\phi_{m}\leq\pi/2$. We compute 
	$p_{ai}=\langle \hat{A}_i \rangle=\Tr(\rho_s\hat{a}_i\hat{a}_i^\dagger) $, $p_{bi}=\langle \hat{B}_i \rangle=\Tr(\rho_s\hat{b}_i\hat{b}_i^\dagger)$, $p_{ci}=\langle \hat{C}_i \rangle=\Tr((\rho_s\otimes\rho_m )\hat{U}^\dagger(\hat{I} \otimes \hat{a}_i\hat{a}_i^\dagger) \hat{U})$ and $p_{di}=\langle \hat{D}_i \rangle=\Tr((\rho_s\otimes\rho_m )\hat{U}^\dagger(\hat{b}_i\hat{b}_i^\dagger\otimes \hat{I} ) \hat{U})$, which are given as
	\beq
	\begin{aligned}
		\centering
		p_{a1}&=f_1^2, &\\ p_{a2}&=f_2^2, &\\ p_{a3}&=f_3^2,&\\
		p_{b1}&=\frac{1}{4}\{{\left|f_1 \right|}^2+2{\left|f_2\right|}^2+{\left|f_3\right|}^2+(f_1\, f_3^*+f_1^*\, f_3)+\sqrt{2}(f_1\, f_2^*+f_1^*\, f_2)+\sqrt{2}(f_2\, f_3^*+f_2^*\, f_3 )\}, &\\
		p_{b2}&=\frac{1}{2}({\left|f_1 \right|}^2+{\left|f_3 \right|}^2)-(f_1\, f_3^*+f_1^*\, f_3 )&\\
		p_{b3}&=\frac{1}{4}\{{\left|f_1 \right|}^2+2{\left|f_2\right|}^2+{\left|f_3\right|}^2+(f_1\, f_3^*+f_1^*\, f_3 )-\sqrt{2}(f_1\, f_2^*+ f_1^*\, f_2 )-\sqrt{2}(f_2\, f_3^*+ f_2^*\, f_3 )\}, &\\
		p_{c1}&={\left|{f_1}\right|}^2\, {\cos^2\!{\phi_m}}\, {\sin^2{\theta}} + {|f_2|}^2\, {\cos^2\!{\theta}} + {|f_3|}^2\, {\sin^2\!\mathrm{\phi_m}}\, {\sin^2\!\mathrm{\theta}}	,&\\
		p_{c2}&={\left|{f_1}\right|}^2\, {\sin^2\!{\phi_m}}\, {\sin^2\!{\theta}} + {|f_2|}^2\, {\cos^2\!{\phi_m}}\, {\sin^2\!{\theta}} + {|f_3|}^2\,{\cos^2\!{\theta}} ,&\\
		p_{c3}&={\left|{f_1}\right|}^2\,{\cos^2\!{\theta}} +{|f_2|}^2\,{\sin^2\!{\phi_m}}\, {\sin^2\!{\theta}} +{|f_3|}^2\,{\cos^2\!{\phi_m}}\, {\sin^2\!{\theta}},&\\
		p_{d1}&=\frac{1}{4}({\left|f_1\right|}^2+2{\left|f_2\right|}^2+{\left|f_3\right|}^2)+\frac{1}{8}\{\sin{2\phi_m}\sin^2{\theta}+\sin{2\theta}(\cos{\phi_m}+\sin{\phi_m})\}&\\
		&\qquad\{(f_1\, f_3^*+f_1^*\, f_3 )+\sqrt{2}(f_1\, f_2^*+f_1^*\, f_2 )+\sqrt{2}(f_2\, f_3^*+f_2^*\, f_3 )\},&\\
		p_{d2}&=\frac{1}{2}({\left|f_1 \right|}^2+{\left|f_3\right|}^2)-\frac{1}{4}\{\sin{2\phi_m}\sin^2{\theta}+\sin{2\theta}(\cos{\phi_m}+\sin{\phi_m})\}&\\
		&\qquad \{(f_1\, f_3^*+f_1^*\, f_3 )\},&\\	
		p_{d3}&=\frac{1}{4}({\left|f_1 \right|}^2+2{\left|f_2\right|}^2+{\left|f_3\right|}^2)+\frac{1}{8}\{\sin{2\phi_m}\sin^2{\theta}+\sin{2\theta}(\cos{\phi_m}+\sin{\phi_m})\}&\\
		&\qquad\{(f_1\, f_3^*+f_1^*\, f_3 )-\sqrt{2}(f_1\, f_2^*+f_1^*\, f_2 )-\sqrt{2}(f_2\, f_3^*+f_2^*\, f_3)\}.&\\
		\centering
	\end{aligned}
	\eeq
	\begin{figure}[htbp]
		\centering
		\includegraphics[width=16cm]{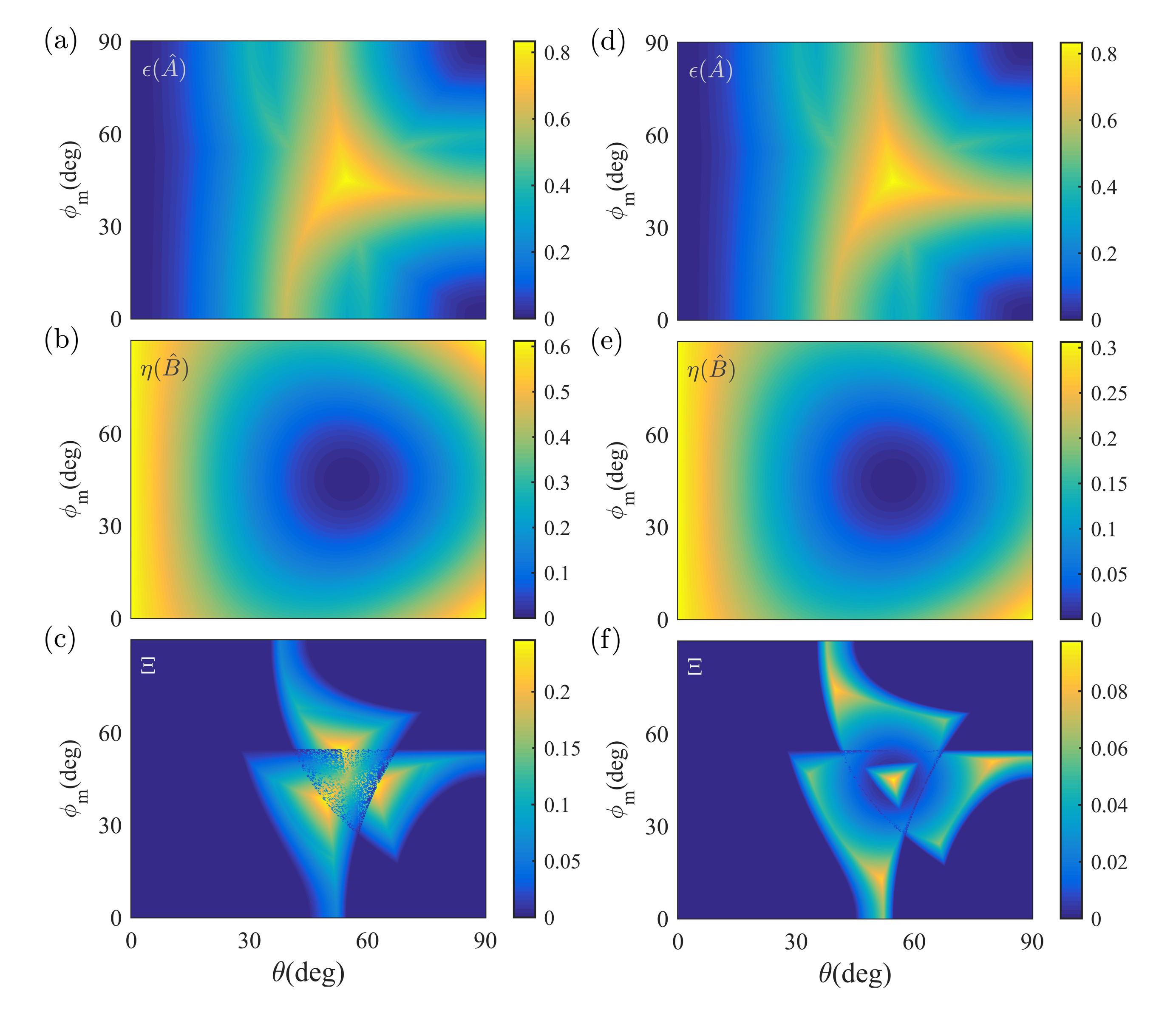}
		\caption{Computed error (a), disturbance (b), and $\Xi$ (c) for qutrits under a number of experimental conditions, with amplitude indicated by the color bar. We fix the system state with $\alpha=\pi/3$, $\phi_s=\pi/2$, and $\chi_{12}=\chi_{13}=0$ ($\chi_{12}=\pi/6$, $\chi_{13}=\pi/3$) in (a)--(c) ((d)--(f)). $\theta \in[0,\pi/2]$ and $\phi_m \in[0,\pi/2]$.}
		\label{fig:FIGS2}
	\end{figure}
	Note that for $\theta=\pi/2$ and $\phi_m=0$, we have $p_{ci}=p_{ai}$, the measurement of $\hat{C}$ is identical to the measurement of $\hat{A}$ , which is the projective measurement performed on the meter state. For $\theta=\arctan(\sqrt{2}),\phi_m=\pi/4$, we have $p_{di}=p_{bi}$, which is the weak measurement limit in this case.
	
	For a given system state, with $\alpha=\pi/3$, $\phi_s=\pi/2$, and $\chi_{12}=\chi_{13}=0$ ($\chi_{12}=\pi/6$, $\chi_{13}=\pi/3$), we plot the computed error, disturbance and $\Xi$ for the measurements of $\hat{A}=\hat{L}_z$ and $\hat{B}=\hat{L}_x$ in Fig. \ref{fig:FIGS2}(a)--(c) (\ref{fig:FIGS2}(d)--(f)), respectively. It is evident that $\Xi\geq0$, hence confirming relation (4).
	
	\section{Calculation of the probabilities of four measurements: $\hat{A}$, $\hat{B}$, $\hat{C}$ and $\hat{D}$ for qubit state}
\begin{figure}[htbp]
		\centering
		\includegraphics[width=10cm]{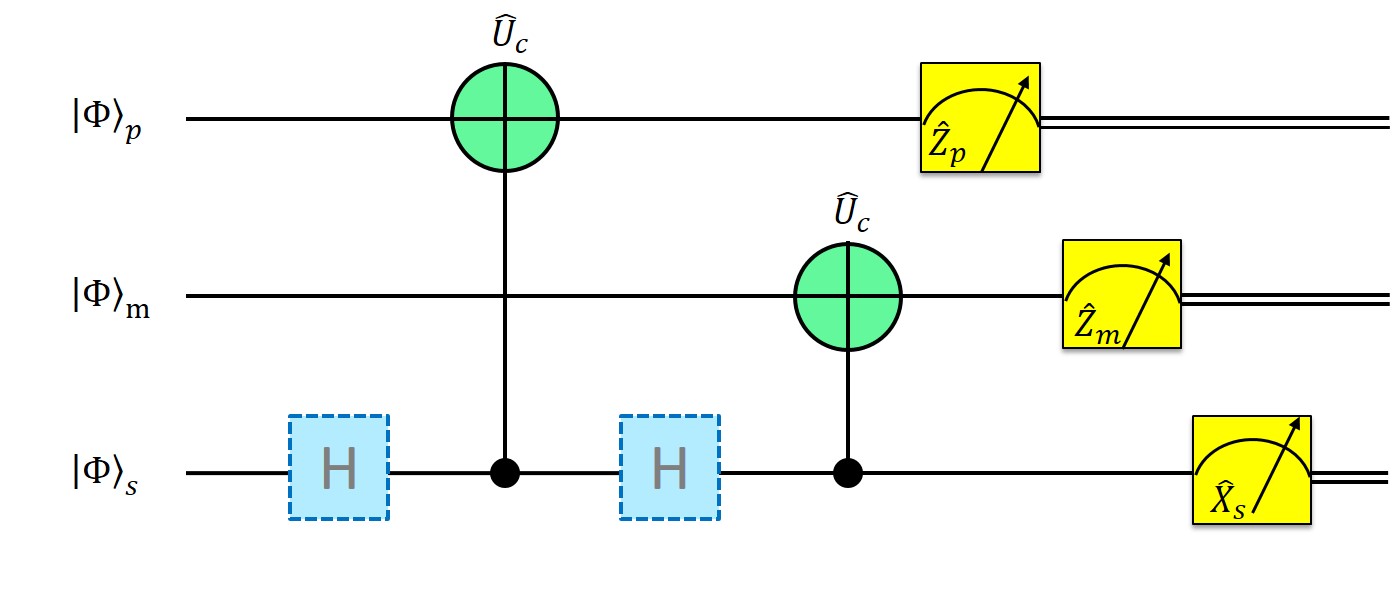}
		\caption{ Quantum circuit model of measuring observable $\hat{X}$ ($\hat{Z}$) with (without) applying Hadamard gates H in a 3-qubit system. As shown in the main text, The system state is sequentially coupled to the probe qubit and the meter qubit by two CNOT gates, with the measurement outcomes of $z_p$, $z_m$ and $x_s$, respectively.}
		\label{fig:QC}
	\end{figure}

	In this section, we detailedly analyze the quantum circuit model \cite{Lund,PhysRevLett.114.070402} of measuring $\hat{A}=\hat{Z}$ and $\hat{B}=\hat{X}$ in Fig. \ref{fig:QC} (Fig. 1 (a) in the main text). The top and middle wires represent the probe state $\ket{\Phi}_p=\gamma\ket{0}_p+\bar{\gamma}\ket{1}_p$ and the meter state $\ket{\Phi}_{m}=\cos\theta\ket{0}_{m}+\sin\theta\ket{1}_{m}$, while the bottom wire corresponds to the system state $\ket{\Phi}_s=\delta\ket{0}_s+\omega\ket{1}_s$, where, for convenience, we replace $\cos{\alpha}$ and $e^{i\phi}\sin\alpha$ in the main text with $\delta$ and $\omega$, respectively. Each state is in a 2-dimensional Hilbert spaces $\mathcal{H}_s$, $\mathcal{H}_p$ and $\mathcal{H}_m$, respectively. It is obvious that we can prepare $\ket{\Phi}_{s}\otimes\ket{\Phi}_p\otimes\ket{\Phi}_m$ in Hilbert space $\mathcal{H}_s\otimes\mathcal{H}_p\otimes\mathcal{H}_m$ as the input state. All three states are properly normalized. The system state is sequentially coupled to the probe qubit and the meter qubit by two CNOT gates. Two Hadamard gates $H$ are inserted to the system state before and after the first CNOT gate when the weak measurement for $\hat{X}$ is taken. The projective measurement $\hat{Z}$, $\hat{Z}$, $\hat{X}$ is separately performed on the probe state, meter state and system state, with the corresponding measurement outcomes represented by $z_p$, $z_m$ and $x_s$, respectively. Then, the outcomes of our scheme can be described as joint measurement of three $\pm1$ three valued observables, with probabilities
	\beq
	\begin{aligned}
		P_{jkl}=P(z_p=j,z_m=k,x_s=l) \quad with \quad j,k,l\in\pm,
	\end{aligned}
	\eeq
	which are determined by a set of positive-operator-valued-measures (POVMs) $\{\Pi_{jkl}\}$. Associated with the POVMs, it is straightforward to calculate the probabilities $\{p_{a\pm}\}$, $\{p_{b\pm}\}$, $\{p_{c\pm}\}$ and $\{p_{d\pm}\}$ of the measurements $\hat{A}$, $\hat{B}$, $\hat{C}$, $\hat{D}$, respectively. Next, we will show the calculation of $\{p_{a\pm}\}$, $\{p_{b\pm}\}$, $\{p_{c\pm}\}$ and $\{p_{d\pm}\}$ in detail.
	
	\subsection{Calculation of the probabilities $\{p_{a\pm}\}$ and $\{p_{c\pm}\}$ for qubit state}
	Following the procedures in Ref \cite{PhysRevLett.114.070402}, after coupling the input system state to the probe state by the first CNOT gate, the state is given as
	\beq
	\begin{aligned}
		\ket{\psi}_1&=CNOT\ket{\Phi}_{p}\otimes\ket{\Phi}_s&\\
		&=CNOT(\gamma\ket{0}_{p}+\bar{\gamma}\ket{1}_{p})\otimes(\delta\ket{0}_s+\omega\ket{1}_s)&\\
		&=(\delta\gamma\ket{0}_{p}+\delta\bar{\gamma}\ket{1}_{p})\otimes\ket{0}_s+(\omega\gamma\ket{1}_{p}+\omega\bar{\gamma}\ket{0}_{p})\otimes\ket{1}_s&\\
		&=\ket{p_0}\otimes\ket{0}_s+\ket{p_1}\otimes\ket{1}_s,&\\
	\end{aligned}
	\eeq
	where
	\beq
	\begin{aligned}
		\ket{p_0}&=\delta\gamma\ket{0}_{p}+\delta\bar{\gamma}\ket{1}_{p},&
		\ket{p_1}&=\omega\gamma\ket{1}_{p}+\omega\bar{\gamma}\ket{0}_{p}.&
	\end{aligned}
	\eeq	
	
	Then, the meter state is added into the whole system through the second CNOT gate. Hence the final system state $\ket{\psi}_f$ is evolved to be
	\begin{equation*}
	\begin{aligned}
	\ket{\psi}_f&=CNOT(\ket{p_0}\otimes\ket{0}_s+\ket{p_1}\otimes\ket{1}_s)\otimes(\cos\theta\ket{0}_m+\sin\theta\ket{1}_m)&\\
	&=\ket{p_0}\otimes\ket{0}_s\otimes(\cos\theta\ket{0}_m+\sin\theta\ket{1}_m)+\ket{p_1}\otimes\ket{1}_s\otimes(\cos\theta\ket{1}_m+\sin\theta\ket{0}_m)&\\
	&=\ket{p_0}\otimes\ket{0}_s\otimes\ket{m_0}+\ket{p_1}\otimes\ket{1}_s\otimes\ket{m_1},&
	\end{aligned}
	\end{equation*}
	where
	\beq
	\begin{aligned}
		\ket{m_0}=\cos\theta\ket{0}_m+\sin\theta\ket{1}_m,\\
		\ket{m_1}=\cos\theta\ket{1}_m+\sin\theta\ket{0}_m.
	\end{aligned}
	\eeq
	Denote $\ket{+}_s=\frac{1}{\sqrt{2}}(\ket{0}_s+\ket{1}_s)$ and $\ket{-}_s=\frac{1}{\sqrt{2}}(\ket{0}_s-\ket{1}_s)$, which are the eigenstates of $\hat{X}$. $\ket{\psi}_f$ can be rewritten as
	\begin{equation*}
	\begin{aligned}
	\ket{\psi}_f&=\ket{p_0}\otimes\ket{0}_s\otimes\ket{m_0}+\ket{p_1}\otimes\ket{1}_s\otimes\ket{m_1}&\\
	&=\frac{1}{\sqrt{2}}(\ket{p_0}\otimes\ket{+}_s\otimes\ket{m_0}+\ket{p_0}\otimes\ket{-}_s\otimes\ket{m_0}+\ket{p_1}\otimes\ket{+}_s\otimes\ket{m_1}-\ket{p_1}\otimes\ket{-}_s\otimes\ket{m_1})&\\
	&=\frac{1}{\sqrt{2}}[(\delta\gamma\cos{\theta}+\omega\bar{\gamma}\sin{\theta})\ket{0}_p\otimes\ket{+}_s\otimes\ket{0}_m
	+(\delta\gamma\sin{\theta}+\omega\bar{\gamma}\cos{\theta})\ket{0}_p\otimes\ket{+}_s\otimes\ket{1}_m&\\
	&+(\delta\bar\gamma\cos{\theta}+\omega{\gamma}\sin{\theta})\ket{1}_p\otimes\ket{+}_s\otimes\ket{0}_m
	+(\delta\bar\gamma\sin{\theta}+\omega{\gamma}\cos{\theta})\ket{1}_p\otimes\ket{+}_s\otimes\ket{1}_m&\\
	&+(\delta\gamma\cos{\theta}-\omega\bar{\gamma}\sin{\theta})\ket{0}_p\otimes\ket{-}_s\otimes\ket{0}_m
	+(\delta\gamma\sin{\theta}-\omega\bar{\gamma}\cos{\theta})\ket{0}_p\otimes\ket{-}_s\otimes\ket{1}_m&\\
	&+(\delta\bar\gamma\cos{\theta}-\omega{\gamma}\sin{\theta})\ket{1}_p\otimes\ket{-}_s\otimes\ket{0}_m
	+(\delta\bar\gamma\sin{\theta}-\omega{\gamma}\cos{\theta})\ket{1}_p\otimes\ket{-}_s\otimes\ket{1}_m].
	\end{aligned}
	\end{equation*}
	
	Finally, we can perform the projective measurement $\hat{Z}$, $\hat{X}$ and $\hat{Z}$ on the probe state, system state and meter state, respectively. Hence, the probabilities of each outcome can be read out as $P_{jkl}$=$P(z_p=j,z_m=k,x_s=l)$:
	\beq
	\begin{aligned}
		8P_{+++}=4|(\delta\gamma\cos{\theta}+\omega\bar{\gamma}\sin{\theta})|^2=1+(2\gamma^2-1)\cos2\theta+(2\gamma^2-1+\cos2\theta)(|\delta|^2-|\omega|^2)+2\gamma\bar\gamma\sin2\theta(\delta^*\omega+\delta\omega^*),\\
		8P_{++-}=4|(\delta\gamma\sin{\theta}+\omega\bar{\gamma}\cos{\theta})|^2=1-(2\gamma^2-1)\cos2\theta+(2\gamma^2-1-\cos2\theta)(|\delta|^2-|\omega|^2)+2\gamma\bar\gamma\sin2\theta(\delta^*\omega+\delta\omega^*),\\
		8P_{-++}=4|(\delta\bar\gamma\cos{\theta}+\omega{\gamma}\sin{\theta})|^2=1-(2\gamma^2-1)\cos2\theta-(2\gamma^2-1-\cos2\theta)(|\delta|^2-|\omega|^2)+2\gamma\bar\gamma\sin2\theta(\delta^*\omega+\delta\omega^*),\\
		8P_{-+-}=4|(\delta\bar\gamma\sin{\theta}+\omega{\gamma}\cos{\theta})|^2=1+(2\gamma^2-1)\cos2\theta-(2\gamma^2-1+\cos2\theta)(|\delta|^2-|\omega|^2)+2\gamma\bar\gamma\sin2\theta(\delta^*\omega+\delta\omega^*),\\
		8P_{+-+}=4|(\delta\gamma\cos{\theta}+\omega\bar{\gamma}\sin{\theta})|^2=1+(2\gamma^2-1)\cos2\theta+(2\gamma^2-1+\cos2\theta)(|\delta|^2-|\omega|^2)-2\gamma\bar\gamma\sin2\theta(\delta^*\omega+\delta\omega^*),\\
		8P_{+--}=4|(\delta\gamma\sin{\theta}+\omega\bar{\gamma}\cos{\theta})|^2=1-(2\gamma^2-1)\cos2\theta+(2\gamma^2-1-\cos2\theta)(|\delta|^2-|\omega|^2)-2\gamma\bar\gamma\sin2\theta(\delta^*\omega+\delta\omega^*),\\
		8P_{--+}=4|(\delta\bar\gamma\cos{\theta}+\omega{\gamma}\sin{\theta})|^2=1-(2\gamma^2-1)\cos2\theta-(2\gamma^2-1-\cos2\theta)(|\delta|^2-|\omega|^2)-2\gamma\bar\gamma\sin2\theta(\delta^*\omega+\delta\omega^*),\\
		8P_{---}=4|(\delta\bar\gamma\sin{\theta}+\omega{\gamma}\cos{\theta})|^2=1+(2\gamma^2-1)\cos2\theta-(2\gamma^2-1+\cos2\theta)(|\delta|^2-|\omega|^2)-2\gamma\bar\gamma\sin2\theta(\delta^*\omega+\delta\omega^*).\\
	\end{aligned}
	\eeq
	This gives the respective 8-outcome POVMs with positive operators $\hat{\Pi}_{jkl} $ on the target system:
	\beq
	\begin{aligned}
		8\hat{\Pi}_{+++}=(1+(2\gamma^2-1)\cos2\theta)\hat{I}+(2\gamma^2-1+\cos2\theta)\hat{Z}+2\gamma\bar\gamma\sin2\theta \hat{X},\\
		8\hat{\Pi}_{++-}=(1-(2\gamma^2-1)\cos2\theta)\hat{I}+(2\gamma^2-1-\cos2\theta)\hat{Z}+2\gamma\bar\gamma\sin2\theta \hat{X},\\
		8\hat{\Pi}_{-++}=(1-(2\gamma^2-1)\cos2\theta)\hat{I}-(2\gamma^2-1-\cos2\theta)\hat{Z}+2\gamma\bar\gamma\sin2\theta \hat{X},\\
		8\hat{\Pi}_{-+-}=(1+(2\gamma^2-1)\cos2\theta)\hat{I}-(2\gamma^2-1+\cos2\theta)\hat{Z}+2\gamma\bar\gamma\sin2\theta \hat{X},\\
		8\hat{\Pi}_{+-+}=(1+(2\gamma^2-1)\cos2\theta)\hat{I}+(2\gamma^2-1+\cos2\theta)\hat{Z}-2\gamma\bar\gamma\sin2\theta \hat{X},\\
		8\hat{\Pi}_{+--}=(1-(2\gamma^2-1)\cos2\theta)\hat{I}+(2\gamma^2-1-\cos2\theta)\hat{Z}-2\gamma\bar\gamma\sin2\theta \hat{X},\\
		8\hat{\Pi}_{--+}=(1-(2\gamma^2-1)\cos2\theta)\hat{I}-(2\gamma^2-1-\cos2\theta)\hat{Z}-2\gamma\bar\gamma\sin2\theta \hat{X},\\
		8\hat{\Pi}_{---}=(1+(2\gamma^2-1)\cos2\theta)\hat{I}-(2\gamma^2-1+\cos2\theta)\hat{Z}-2\gamma\bar\gamma\sin2\theta \hat{X}.\\
	\end{aligned}
	\eeq
	According to the definition of the POVMs, $\hat{\Pi}_{j=\pm}=\sum_{kl}{\hat{\Pi}_{jkl}}$ represents the initial weak $\hat{Z}$ measurement:
	\beq
	\begin{aligned}
		\hat{\Pi}_{j=\pm}=\frac{1}{2}(\hat{I}\pm(2\gamma^2-1)\hat{Z}).
	\end{aligned}
	\eeq
	We take advantage of the outcomes of the weak measurement to calculate the POVM of $\hat{A}$,
	\beq
	\begin{aligned}
		\hat{\Pi}_{j=\pm}^a=\frac{1}{2}(1\pm\frac{\hat{\Pi}_{j=+}-\hat{\Pi}_{j=-}}{2\gamma^2-1}).\\
	\end{aligned}
	\eeq
	Correspondingly, the probabilities of measuring observable $\hat{A}$ are given by
	\beq
	\begin{aligned}
		p_{a\pm}=\frac{1}{2}(1\pm\frac{\langle\hat{\Pi}_{j=+}\rangle-\langle\hat{\Pi}_{j=-}\rangle}{2\gamma^2-1}).
		\label{eq:PA}
	\end{aligned}
	\eeq
	Following the similar procedure, the POVM of $\hat{C}$, which is the general measurement associated with the observable $\hat{A}$, can be calculated,
	\beq
	\begin{aligned}
		\hat{C}_{\pm}=\hat{\Pi}_{k=\pm}=\sum_{jl}{\hat{\Pi}_{jkl}}=\frac{1}{2}(\hat{I}\pm\cos2\theta \hat{Z}).\\
	\end{aligned}
	\eeq
	The probabilities with respect to $\hat{C}$ are:
	\beq
	\begin{aligned}
		p_{c\pm}=\langle\hat{\Pi}_{k=\pm}\rangle=\frac{1}{2}(1\pm\cos2\theta \langle\hat{Z}\rangle).\\
	\end{aligned}
	\eeq
	
	\subsection{Calculation of the probabilities $\{p_{b\pm}\}$ and $\{p_{d\pm}\}$ for qubit state}
	In order to get the probabilities $\{p_{b\pm}\}$ and $\{p_{d\pm}\}$, two Hadamard gates $H$ are inserted to the system state before and after the first CNOT gate. Everything else in the model is entirely identical as before. We still continue to perform similar steps.
	Firstly, the state $\ket{\psi}_1$ is given by
	\beq
	\begin{aligned}
		\ket{\psi}_1&=(I\otimes H)CNOT(I\otimes H)\ket{\Phi}_p\otimes\ket{\Phi}_s&\\
		&=(I\otimes H)CNOT(I\otimes H)(\gamma\ket{0}_p+\bar{\gamma}\ket{1}_p)\otimes(\delta\ket{0}_s+\omega\ket{1}_s)&\\
		&=\frac{1}{\sqrt{2}}(I\otimes H)CNOT(\gamma\ket{0}_p+\bar{\gamma}\ket{1}_p)\otimes((\delta+\omega)\ket{0}_s+(\delta-\omega)\ket{1}_s)&\\
		&=\frac{1}{\sqrt{2}}(I\otimes H)(\gamma\ket{0}_p+\bar{\gamma}\ket{1}_p)\otimes((\delta+\omega)\ket{0}_s)+(\gamma\ket{1}_p+\bar{\gamma}\ket{0}_p)\otimes(\delta-\omega)\ket{1}_s)&\\
		&=\frac{1}{2}(\gamma\ket{0}_p+\bar{\gamma}\ket{1}_p)\otimes(\delta+\omega)(\ket{0}_s+\ket{1}_s)+(\gamma\ket{1}_p+\bar{\gamma}_p\ket{0})\otimes(\delta-\omega)(\ket{0}_s-\ket{1}_s))&\\
		&=\frac{1}{2}[(\gamma(\delta+\omega)+\bar{\gamma}(\delta-\omega))\ket{0}_p+(\bar\gamma(\delta+\omega)+\gamma(\delta-\omega))\ket{1}_p]\otimes\ket{0}_s&\\
		&+[(\gamma(\delta+\omega)-\bar{\gamma}(\delta-\omega))\ket{0}_p+(\bar\gamma(\delta+\omega)-\gamma(\delta-\omega))\ket{1}_p]\otimes\ket{1}_s&\\
		&=\ket{p_0}\otimes\ket{0}_s+\ket{p_1}\otimes\ket{1}_s,&\\
	\end{aligned}
	\eeq
	where
	\beq
	\begin{aligned}
		\ket{p_0}&=\frac{1}{2}(\gamma(\delta+\omega)+\bar{\gamma}(\delta-\omega))\ket{0}_p+(\bar\gamma(\delta+\omega)+\gamma(\delta-\omega))\ket{1}_p,&\\
		\ket{p_1}&=\frac{1}{2}(\gamma(\delta+\omega)-\bar{\gamma}(\delta-\omega))\ket{0}_p+(\bar\gamma(\delta+\omega)-\gamma(\delta-\omega))\ket{1}_p.&
	\end{aligned}
	\eeq
	Secondly, the meter state is added into the whole system through the second CNOT gate. The final system state $\ket{\psi}_f$ is evolved to be:
	\beq
	\begin{aligned}
		\ket{\psi}_f&=CNOT(\ket{p_0}\otimes\ket{0}_s+\ket{p_1}\otimes\gamma\ket{1}_s)\otimes(\cos\theta\ket{0}_m+\sin\theta\ket{1}_m)&\\
		&=\ket{p_0}\otimes\ket{0}_s\otimes(\cos\theta\ket{0}_m+\sin\theta\ket{1}_m)+\ket{p_1}\otimes\gamma\ket{1}_s\otimes(\cos\theta\ket{1}_m+\sin\theta\ket{0}_m)&\\
		&=\ket{p_0}\otimes\ket{0}_s\otimes\ket{m_0}+\ket{p_1}\otimes\ket{1}_s\otimes\ket{m_1},&
	\end{aligned}
	\eeq
	where
	\beq
	\begin{aligned}
		\ket{m_0}=\cos\theta\ket{0}_m+\sin\theta\ket{1}_m,\\
		\ket{m_1}=\cos\theta\ket{1}_m+\sin\theta\ket{0}_m.
	\end{aligned}
	\eeq
	Denote $\ket{+}_s=\frac{1}{\sqrt{2}}(\ket{0}_s+\ket{1}_s)$ and $\ket{-}_s=\frac{1}{\sqrt{2}}(\ket{0}_s-\ket{1}_s)$. $\ket{\psi}_f$ can be written as
	\beq
	\begin{aligned}
		\ket{\psi}_f&=\ket{p_0}\otimes\ket{0}_s\otimes\ket{m_0}+\ket{p_1}\otimes\ket{1}_s\otimes\ket{m_1}&\\
		&=\frac{1}{\sqrt{2}}(\ket{p_0}\otimes\ket{+}_s\otimes\ket{m_0}+\ket{p_0}\otimes\ket{-}_s\otimes\ket{m_0}+\ket{p_1}\otimes\ket{+}_s\otimes\ket{m_1}-\ket{p_1}\otimes\ket{-}_s\otimes\ket{m_1})&\\
		&=\frac{1}{2\sqrt{2}}[((\gamma(\delta+\omega)+\bar{\gamma}(\delta-\omega))\cos{\theta}+(\gamma(\delta+\omega)-\bar{\gamma}(\delta-\omega))\sin{\theta})\ket{0}_p\otimes\ket{+}_s\otimes\ket{0}_m&\\
		&+((\gamma(\delta+\omega)+\bar{\gamma}(\delta-\omega))\sin{\theta}+(\gamma(\delta+\omega)-\bar{\gamma}(\delta-\omega))\cos{\theta})\ket{0}_p\otimes\ket{+}_s\otimes\ket{1}_m&\\
		&+((\bar\gamma(\delta+\omega)+\gamma(\delta-\omega))\cos{\theta}+(\bar\gamma(\delta+\omega)-\gamma(\delta-\omega))\sin{\theta})\ket{1}_p\otimes\ket{+}_s\otimes\ket{0}_m&\\
		&+((\bar\gamma(\delta+\omega)+\gamma(\delta-\omega))\sin{\theta}+(\bar\gamma(\delta+\omega)-\gamma(\delta-\omega))\cos{\theta})\ket{1}_p\otimes\ket{+}_s\otimes\ket{1}_m&\\
		&+((\gamma(\delta+\omega)+\bar{\gamma}(\delta-\omega))\cos{\theta}-(\gamma(\delta+\omega)-\bar{\gamma}(\delta-\omega))\sin{\theta})\ket{0}_p\otimes\ket{-}_s\otimes\ket{0}_m&\\
		&+((\gamma(\delta+\omega)+\bar{\gamma}(\delta-\omega))\sin{\theta}-(\gamma(\delta+\omega)-\bar{\gamma}(\delta-\omega))\cos{\theta})\ket{0}_p\otimes\ket{-}_s\otimes\ket{1}_m&\\
		&+((\bar\gamma(\delta+\omega)+\gamma(\delta-\omega))\cos{\theta}-(\bar\gamma(\delta+\omega)-\gamma(\delta-\omega))\sin{\theta})\ket{1}_p\otimes\ket{-}_s\otimes\ket{0}_m&\\
		&+((\bar\gamma(\delta+\omega)+\gamma(\delta-\omega))\sin{\theta}-(\bar\gamma(\delta+\omega)-\gamma(\delta-\omega))\cos{\theta})\ket{1}_p\otimes\ket{-}_s\otimes\ket{1}_m].&\\
	\end{aligned}
	\eeq
	
	Finally, we can carry out the projective measurement $\hat{Z}$, $\hat{X}$ and $\hat{Z}$ on the probe state, system state and meter state, respectively. Hence, the probabilities of each outcome can be read out as $P_{jkl}$:
	
	\beq
	\begin{aligned}
		8P_{+++}=1+(2\gamma^2-1)\sin2\theta+(2\gamma^2-1+\sin2\theta)(\delta^*\omega+\delta\omega^*)+2\gamma\bar\gamma\cos2\theta(|\delta|^2-|\omega|^2),\\
		8P_{++-}=1+(2\gamma^2-1)\sin2\theta+(2\gamma^2-1+\sin2\theta)(\delta^*\omega+\delta\omega^*)-2\gamma\bar\gamma\cos2\theta(|\delta|^2-|\omega|^2),\\
		8P_{-++}=1+(1-2\gamma^2)\sin2\theta+(1-2\gamma^2+\sin2\theta)(\delta^*\omega+\delta\omega^*)+2\gamma\bar\gamma\cos2\theta(|\delta|^2-|\omega|^2),\\
		8P_{-+-}=1+(1-2\gamma^2)\sin2\theta+(1-2\gamma^2+\sin2\theta)(\delta^*\omega+\delta\omega^*)-2\gamma\bar\gamma\cos2\theta(|\delta|^2-|\omega|^2),\\
		8P_{+-+}=1-(2\gamma^2-1)\sin2\theta+(2\gamma^2-1-\sin2\theta)(\delta^*\omega+\delta\omega^*)+2\gamma\bar\gamma\cos2\theta(|\delta|^2-|\omega|^2),\\
		8P_{+--}=1-(2\gamma^2-1)\sin2\theta+(2\gamma^2-1-\sin2\theta)(\delta^*\omega+\delta\omega^*)-2\gamma\bar\gamma\cos2\theta(|\delta|^2-|\omega|^2),\\
		8P_{--+}=1-(1-2\gamma^2)\sin2\theta+(1-2\gamma^2-\sin2\theta)(\delta^*\omega+\delta\omega^*)+2\gamma\bar\gamma\cos2\theta(|\delta|^2-|\omega|^2),\\
		8P_{---}=1-(1-2\gamma^2)\sin2\theta+(1-2\gamma^2-\sin2\theta)(\delta^*\omega+\delta\omega^*)-2\gamma\bar\gamma\cos2\theta(|\delta|^2-|\omega|^2).\\
	\end{aligned}
	\eeq	
	The	$P_{jkl}$ corresponds to 8 POVM $\hat{\Pi}_{jkl} $ on the target system:
	\beq
	\begin{aligned}
		8\hat{\Pi}_{+++}=(1+(2\gamma^2-1)\sin2\theta)\hat{I}+(2\gamma^2-1+\sin2\theta)\hat{X}+2\gamma\bar\gamma\cos2\theta\hat{Z},\\
		8\hat{\Pi}_{++-}=(1+(2\gamma^2-1)\sin2\theta)\hat{I}+(2\gamma^2-1+\sin2\theta)\hat{X}-2\gamma\bar\gamma\cos2\theta\hat{Z},\\
		8\hat{\Pi}_{-++}=(1+(1-2\gamma^2)\sin2\theta)\hat{I}+(1-2\gamma^2+\sin2\theta)\hat{X}+2\gamma\bar\gamma\cos2\theta\hat{Z},\\
		8\hat{\Pi}_{-+-}=(1+(1-2\gamma^2)\sin2\theta)\hat{I}+(1-2\gamma^2+\sin2\theta)\hat{X}-2\gamma\bar\gamma\cos2\theta\hat{Z},\\
		8\hat{\Pi}_{+-+}=(1-(2\gamma^2-1)\sin2\theta)\hat{I}+(2\gamma^2-1-\sin2\theta)\hat{X}+2\gamma\bar\gamma\cos2\theta\hat{Z},\\
		8\hat{\Pi}_{+--}=(1-(2\gamma^2-1)\sin2\theta)\hat{I}+(2\gamma^2-1-\sin2\theta)\hat{X}-2\gamma\bar\gamma\cos2\theta\hat{Z},\\
		8\hat{\Pi}_{--+}=(1-(1-2\gamma^2)\sin2\theta)\hat{I}+(1-2\gamma^2-\sin2\theta)\hat{X}+2\gamma\bar\gamma\cos2\theta\hat{Z},\\
		8\hat{\Pi}_{---}=(1-(1-2\gamma^2)\sin2\theta)\hat{I}+(1-2\gamma^2-\sin2\theta)\hat{X}-2\gamma\bar\gamma\cos2\theta\hat{Z}.\\
	\end{aligned}
	\eeq
	Here $\hat{\Pi}_j=\sum_{kl}{\hat{\Pi}_{jkl}}$ represents the initial weak $\hat{X}$ measurement:
	\beq
	\begin{aligned}
		\hat{\Pi}_{j=\pm}=\frac{1}{2}(\hat{I}\pm(2\gamma^2-1)\hat{X}).\\
	\end{aligned}
	\eeq
	Associated with the POVM $\hat{B}$, one has
	\beq
	\begin{aligned}
		\hat{\Pi}_{j=\pm}^b=\frac{1}{2}(1\pm\frac{\hat{\Pi}_{j=+}-\hat{\Pi}_{j=-}}{2\gamma^2-1}).\\
	\end{aligned}
	\eeq
	The corresponding probabilities with respect to the observable $\hat{B}$ are given by
	\beq
	\begin{aligned}
		p_{b\pm}=\frac{1}{2}(1\pm\frac{\langle\hat{\Pi}_{j=+}\rangle-\langle\hat{\Pi}_{j=-}\rangle}{2\gamma^2-1}).
		\label{eq:PB}
	\end{aligned}
	\eeq
	Similarly, for the POVM $\hat{D}$, which is the general measurement associated with the observable $\hat{B}$, we have
	\beq
	\begin{aligned}
		\hat{D}_{\pm}=\hat{\Pi}_{l=\pm}=\sum_{jk}{\hat{\Pi}_{jkl}}=\frac{1}{2}(\hat{I}\pm\sin2\theta \hat{X}),
	\end{aligned}
	\eeq
	with the corresponding probabilities
	\beq
	\begin{aligned}
		p_{d\pm}=\langle\hat{\Pi}_{l=\pm}\rangle=\frac{1}{2}(1\pm\sin2\theta \langle\hat{X}\rangle).\\
	\end{aligned}
	\eeq
	
	\section{The Sagnac interferometer}
	
	In the Sagnac interferometer in Fig. 1 (b) \cite{PhysRevLett.112.020402}, the polarization and path degree of freedoms of single photons acts as the system qubit and probe (meter) qubit, respectively. The polarization beams plitter (PBS) serves as the CNOT gate controlled by system qubit. The probabilities of joint measurement are strongly correlated with the performance of PBSs in the measurement setup. In order to take account of the unideal extinction ratio of PBS \cite{baek2013experimental}, we define the reflection extinction ratio as follows:
	\beq
	\begin{aligned}
		e_r=\frac{R_V}{R_H},
	\end{aligned}
	\eeq
	where the quantities $R_V$ and $R_H$ are the PBS reflectance of vertical polarization and horizontal polarization, respectively. It's obvious that when $R_V=1, R_H=0$, $e_r=\infty$. We also define $T_V, T_H$ are the PBS transmittance of vertical polarization and horizontal polarization. The PBS behaviour can be accounted by the general unitary matrix:
	\beq
	\begin{aligned}
		\hat{U}_{PBS}=\ket{0}_{pol}{}_{pol}\bra{0}\otimes (\sqrt{T_H}\hat{I}_{path}+i\sqrt{R_H}\hat{X}_{path})+\ket{1}_{pol}{}_{pol}\bra{1}\otimes(i\sqrt{T_V}\hat{I}_{path}+\sqrt{R_V}\hat{X}_{path}).
	\end{aligned}
	\eeq
	We discuss the performence of the Sagnac interferometer with the perfect and imperfect PBSs in sequence.	
	
	\subsection{The Sagnac interferometer with the perfect PBSs} 	
	In the ideal case, the parameters of the PBS satisfy $ R_V = 1$, $ R_H = 0$, $ T_V = 0$ and $ T_H = 1$. Hence the unitary matrix $ \hat{U}_{PBS}=\ket{0}_{pol}{}_{pol}\bra{0}\otimes \hat{I}_{path}+\ket{1}_{pol}{}_{pol}\bra{1}\otimes\hat{X}_{path}$.
	$\hat{U}_{path}$ is a CNOT gate by a half-wave plate (HWP) oriented at $45^o$ with path qubit as control and polarization qubit as target: $\hat{U}_{path}=\hat{I}_{pol}\otimes\ket{0}_{path}{}_{path}\bra{0}+ \hat{X}_{pol}\otimes\ket{1}_{path}{}_{path}\bra{1}$. The $\hat{U}_{phase}$ is the $\hat{Z}$-phase gate induced by the difference between the reflections of $|H\rangle-$ and $|V\rangle-$ polarized photons on mirrors. $\hat{U}_{\gamma}=(\gamma\hat{Z}_{pol}+\bar{\gamma}\hat{X}_{pol})\otimes \hat{I}_{path} $ is used to couple polarization qubit and path qubit with a HWP. Then the matrix of the Sagnac interferometer can be described by
	\beq
	\begin{aligned}
		\hat{U}_{C}&=\hat{U}_{path} \hat{U}_{PBS} \hat{U}_{\gamma} \hat{U}_{phase} \hat{U}_{path} \hat{U}_{PBS}&\\
		&=(\gamma \ket{0}_{pol}{}_{pol}\bra{0}+\bar{\gamma}\ket{1}_{pol}{}_{pol}\bra{1})\otimes (\ket{0}_{path}{}_{path}\bra{0})&\\
		&+(- \bar{\gamma}\ket{0}_{pol}{}_{pol}\bra{1}+\gamma\ket{1}_{pol}{}_{pol}\bra{0})\otimes (\ket{0}_{path}{}_{path}\bra{1})&\\
		&+(\bar{\gamma}\ket{0}_{pol}{}_{pol}\bra{0}+\gamma\ket{1}_{pol}{}_{pol}\bra{1})\otimes (\ket{1}_{path}{}_{path}\bra{0})&\\
		&+(\gamma\ket{0}_{pol}{}_{pol}\bra{1}-\bar{\gamma}\ket{1}_{pol}{}_{pol}\bra{0})\otimes (\ket{1}_{path}{}_{path}\bra{1})&\\
		&=\left(\begin{array}{cccc} \gamma & 0 & 0 & - \bar{\gamma}\\
			\bar{\gamma}\ & 0 & 0 & \gamma\\
			0 & \gamma& \bar{\gamma} & 0\\
			0 & -\bar{\gamma} & \gamma & 0 \end{array}\right).&
	\end{aligned}
	\eeq
	We prepare the input state $\ket{\Psi}_{i}=\ket{\Phi}_{pol}|0\rangle_{path}=(\cos\alpha\ket{0}_{pol}+e^{i\phi}\sin\alpha \ket{1}_{pol})|0\rangle_{path}$. After passing through the Sagnac interferometer, the output state is evolved into $\hat{U}_{C}\ket{\Psi}_{i}=\hat{U}_{path} \hat{U}_{PBS} \hat{U}_{\gamma}\hat{U}_{phase}\hat{U}_{path}\hat{U}_{PBS}\ket{\Psi}_{i}=(\cos\alpha\gamma|0\rangle_{pol}+e^{i\phi} \sin\alpha\bar{\gamma} |1\rangle_{pol})|0\rangle_{path}+(\cos\alpha \bar{\gamma}|0\rangle_{pol}+e^{i\phi}\sin\alpha\gamma |1\rangle_{pol})|1\rangle_{path}$. By introducing the measurement operators $M_{m'}=\bra{m'}U_{Sagnac}\ket{0'}$ with $m'=0,1$, the measurement can be described as
	\beq
	\begin{aligned}
		\hat{U}_{Sagnac}\ket{\Psi}_i=\hat{M_0}\ket{\Phi}_{pol}\ket{0}_{path}+\hat{M_1}\ket{\Phi}_{pol}\ket{1}_{path},\\
	\end{aligned}
	\eeq
	where
	\beq
	\begin{aligned}
		\hat{M_0}={}_{path}\bra{0}  \hat{U}_{Sagnac} \ket{0}_{path}=\gamma\ket{0}_{pol}{}_{pol}\bra{0}+\bar{\gamma}\ket{1}_{pol}{}_{pol}\bra{1},\\
		\hat{M_1}={}_{path}\bra{1}  \hat{U}_{Sagnac} \ket{0}_{path}=\bar{\gamma}\ket{0}_{pol}{}_{pol}\bra{0}+\gamma\ket{1}_{pol}{}_{pol}\bra{1}.\\
	\end{aligned}
	\eeq
	The corresponding POVM elements are
	\beq
	\begin{aligned}
		\hat{\Pi}_{z=+1}=\hat{M_0}^{+}\hat{M_0}=\frac{1}{2}(\hat{I}+({2\gamma^2-1}) \hat{Z}),\\
		\hat{\Pi}_{z=-1}=\hat{M_1}^{+}\hat{M_1}=\frac{1}{2}(\hat{I}-({2\gamma^2-1}) \hat{Z}).
		\label{eq:POVMA}
	\end{aligned}
	\eeq
	
	For two Sagnac interferometers after the weak measurement, we replace  $\cos\theta$ with $\gamma$, $\sin\theta$ with $\bar{\gamma}$. It's similar to obtain that $\hat{U}_{C}\ket{\Psi}_{i}=\hat{U}_{path} \hat{U}_{PBS} \hat{U}_{\theta} \hat{U}_{phase} \hat{U}_{path} \hat{U}_{PBS} \ket{\Psi}_{i}  =(\cos\alpha\cos\theta  |0\rangle_{pol}+e^{i\phi} \sin\alpha\sin\theta|1\rangle_{pol})|0\rangle_{path}+(\cos\alpha\sin\theta |0\rangle_{pol}+e^{i\phi}\sin\alpha \cos\theta  |1\rangle_{pol})|1\rangle_{path}$. The corresponding POVMs are $\hat{\Pi}_{z=\pm1}=\frac{1}{2}(\hat{I}\pm\cos2\theta\hat{Z})$.
	
	\subsection{The Saganc interferometer with the imperfect PBSs}
	In the real experiment, all the instruments are not perfect. Therefore, it is necessary to consider the POVMs of the Sagnac interferometer with the imperfect extinction ratio of PBSs. Suppose that the extinction ratio of two sides of the PBS diagonal plane is $e_1$ and $e_2$, the unitary operator $\hat{U}_{C}$ of the quantum circuit in Fig. 2 (c) is given by
	\beq
	\begin{aligned}
		\centering
		\hat{U}_{C}=\hat{U}_{path} \hat{U}_{PBS} \hat{U}_{\gamma} \hat{U}_{phase} \hat{U}_{path} \hat{U}_{PBS}=
		\left(\begin{array}{cccc} t_1 t_2 \gamma+ s_1 s_2 \bar{\gamma} & (s_1 t_2 \gamma -t_1 s_2 \bar{\gamma})i
			& s_2 \gamma\,i & - t_{2}\, \bar{\gamma}\\	 t_{1}\, \bar{\gamma} & s_1\gamma\,i & 0 & \gamma\\ s_1\gamma\,i & t_{1}\, \gamma& \bar{\gamma} & 0\\ (t_1 s_2 \gamma -s_1 t_2 \bar{\gamma})i &  - t_{1} t_{2}\, \bar{\gamma}-s_1 s_2\gamma  & t_{2}\, \gamma & -s_2\bar{\gamma} \end{array}\right),
		\centering
	\end{aligned}
	\eeq
	where $s_1=\sqrt{\frac{1}{e_1}}$, $s_2=\sqrt{\frac{1}{e_2}}$, $t_1=\sqrt{{1-\frac{1}{e_1}}}$ and $t_2=\sqrt{{1-\frac{1}{e_2}}}$. 
	Similarly,the measurement operators are given by
	\beq
	\begin{aligned}
		\centering
		\hat{M}_0=\left(\begin{array}{cc} t_{1}\, t_{2}\, \gamma + s_{1}\, s_{2}\, \bar{\gamma}  &  s_2\,\gamma i\\ s_1\,\gamma i& \bar{\gamma} \\ \end{array}\right),
		\centering
	\end{aligned}
	\eeq
	\beq
	\begin{aligned}
		\hat{M}_1=\left(\begin{array}{cccc}  t_{1}\, \bar{\gamma} & 0 \\ (t_1 s_2 \gamma -s_1 t_2 \bar{\gamma})i& t_{2}\, \gamma\end{array}\right).
	\end{aligned}
	\eeq
	The corresponding POVM elements are
	\beq
	\begin{aligned}
		\hat{\Pi}_{z=+1}=\hat{M}_0^{+}\hat{M}_0=a_0 \hat{I} +b_0\hat{X}+c_0\hat{Y}+d_0\hat{Z}, \\
		\hat{\Pi}_{z=-1}=\hat{M}_1^{+}\hat{M}_1=a_1 \hat{I} +b_1 \hat{X}+c_1 \hat{Y}+d_1\hat{Z},
		\label{eq:POVMB}
	\end{aligned}
	\eeq
	where
	\beq
	\begin{aligned}
		\centering
		a&=\frac{1}{2}\left(\begin{array}{c} 1+ 2\gamma\bar{\gamma}\,s_1s_2t_1t_2 +s_1^2s_2^2\\ 1-2\gamma\bar{\gamma}\,s_1s_2t_1t_2  -s_1^2s_2^2 \end{array}\right),&\\
		b&=\left(\begin{array}{c} 0\\0 \end{array}\right),	&	\\
		c&=\frac{1}{2}\left[2\gamma\bar{\gamma}\,s_1t_2^2 -\, {2\gamma^2} t_1s_2t_2\right]\left(\begin{array}{c} 1 \\ -1\end{array}\right),&\\	
		d&=\frac{1}{2}\left[ 2\gamma\bar{\gamma}\, s_1s_2t_1t_2+  ({2\gamma^2-1})\,t_2^2\, - s_2^2 + s_1^2s_2^2 \right]\left(\begin{array}{c} 1\\-1 \end{array}\right).&		\\
		\label{eq:CEO1}
		\centering
	\end{aligned}
	\eeq		
	Expressing $s_1, s_2, t_1, t_2$ in terms of the parameters $e_1,e_2$, we have
	\beq
	\begin{aligned}
		\centering
		a&=\frac{1}{2}\left(\begin{array}{c} 1+2\gamma\bar{\gamma}\,\frac{\sqrt{e_{1} - 1}\, \sqrt{e_{2} - 1}}{e_{1}\, e_{2}}  + \frac{1}{e_{1}\, e_{2}}\\ 1-2\gamma\bar{\gamma} \,\frac{\sqrt{e_{1} - 1}\, \sqrt{e_{2} - 1}}{e_{1}\, e_{2}} - \frac{1}{e_{1}\, e_{2}} \end{array}\right),	&\\
		b&=\left(\begin{array}{c} 0\\0 \end{array}\right),&		\\
		c&=\frac{1}{2}\left[2\gamma\bar{\gamma}\,\left(\frac{1}{\sqrt{e_{1}}} - \frac{1}{\sqrt{e_{1}} e_{2}}\right)\, - {2\gamma^2}\,  \frac{\sqrt{e_{1} - 1}\, \sqrt{e_{2} - 1}}{\sqrt{e_{1}}\, e_{2}} \right]\left(\begin{array}{c} 1 \\ -1\end{array}\right),&\\	
		d&=\frac{1}{2}\left[ 2\gamma\bar{\gamma}\,\frac{\sqrt{e_{1} - 1}\, \sqrt{e_{2} - 1}}{e_{1}\, e_{2}}  + ({2\gamma^2-1})\left(1 - \frac{1}{e_{2}}\right)\,  - \frac{1}{e_{2}} + \frac{1}{e_{1}\, e_{2}} \right]\left(\begin{array}{c} 1\\-1 \end{array}\right).&		\\	
		\label{eq:CEO2}
	\end{aligned}
	\eeq
	It is obvious that with the perfect PBS, there are $e_1=\infty$, and $e_2=\infty$. Hence the coefficients of Pauli matrices are		
	\beq
	\begin{aligned}
		a=\frac{1}{2}\left(\begin{array}{c} 1\\ 1\end{array}\right),	b=\left(\begin{array}{c} 0\\0 \end{array}\right),		c=\left(\begin{array}{c} 0\\0 \end{array}\right),	
		d=\frac{1}{2} ({2\gamma^2-1})\left(\begin{array}{c} 1\\-1 \end{array}\right).	
	\end{aligned}
	\eeq
	The POVMs in Eq. (\ref{eq:POVMB}) reduce to that of Eq. (\ref{eq:POVMA}).
	
	\section{Analysis of the experimental results for the linearly polarized system qubit}
	
	In the section \textbf{\RNum{2}}, the probabilities of $\hat{A}$ is given by $p_{a\pm}=\frac{1}{2}(1\pm\frac{\langle\hat{\Pi}_{j=+}\rangle-\langle\hat{\Pi}_{j=-}\rangle}{2\gamma^2-1})$ in the Eq. (\ref{eq:PA}), where we consider the ideal CNOT gate. For the linearly polarized system qubit, $\ket{\Phi}_{s}=\cos\alpha\ket{0}_{pol}+\sin\alpha\ket{1}_{pol}$, it is obvious that the LHS and RHS of the relation (4) coincide, and they are symmetrical along the straight line $\alpha=45^{\circ}$ in the ideal case, which is shown in the Fig. 3 of the main text. But they are no longer symmetrical in the practical experiment. There is a gap near $\alpha=45^{\circ}$ between red and the blue dots, and the gap becomes smaller as the measured strength becomes smaller (from figure (a) to (e)). It can also be seen that the experimental results are reconstructed by the dashed line when we take the imperfect PBS into account. Therefore the imperfect PBS is the most important factor influencing the results of the experiment. Below we analyze how it affects the results in detail.
	
	In the last section, we have discussed the Sagnac interferometer with the imperfect PBSs in the subsection \textbf{B}, where we calculated the POVMs in Eq. (\ref{eq:POVMB}). If we use these POVMs to calculate the probabilities $p_{a\pm}$, we can find that
	\beq
	\begin{aligned}
		p_{a\pm}&=\frac{1}{2}(1\pm\frac{\langle\hat{\Pi}_{j=+}\rangle-\langle\hat{\Pi}_{j=-}\rangle}{2\gamma^2-1})&\\
		&=\frac{1}{2}(1\pm\frac{(a_0-a_1)+(b_0-b_1)\langle\hat{X}\rangle+(c_0-c_1)\langle\hat{Y}\rangle+(d_0-d_1)\langle\hat{Z}\rangle}{2\gamma^2-1}).\\
	\end{aligned}
	\eeq
	For the linearly polarized system qubit, we have $\langle\hat{X}\rangle=\sin2\alpha$, $\langle\hat{Y}\rangle=0$, $\langle\hat{Z}\rangle=\cos2\alpha$. Combining the coefficients $a, b, c$ and $d$ in the Eq. (\ref{eq:CEO1}) or (\ref{eq:CEO2}), the probabilities can be simplified to
	\beq
	\begin{aligned}
		p_{a\pm}&=\frac{1}{2}(1\pm(\delta_a^{'} +\delta_d^{'} \cos2\alpha)),
		\label{eq:pa}
	\end{aligned}
	\eeq
	where \beq
	\begin{aligned}
		\delta_a^{'}&=\frac{a_0-a_1}{{2\gamma^2-1}}=\frac{2\gamma\bar{\gamma}}{{2\gamma^2-1}}\,\frac{\sqrt{e_{1} - 1}\, \sqrt{e_{2} - 1}}{e_{1}\, e_{2}}  + \frac{1}{{2\gamma^2-1}}\frac{1}{e_{1}\, e_{2}},&\\
		\delta_d^{'}&=\frac{d_0-d_1}{{2\gamma^2-1}}= \frac{2\gamma\bar{\gamma}}{{2\gamma^2-1}}\,\frac{\left(\sqrt{e_{1} - 1}\, \sqrt{e_{2} - 1}\right)}{e_{1}\, e_{2}}  + \left(1 - \frac{1}{e_{2}}\right)\,  +\frac{1}{{2\gamma^2-1}}(\frac{1}{e_{1}\, e_{2}}- \frac{1}{e_{2}}).&	
	\end{aligned}
	\eeq
	In a similar way, we can calculate the probability $p_{b\pm}$ by adding Hardmard gate before and after the Sagnac interferometer.
	\beq
	\begin{aligned}
		p_{b\pm}&=\frac{1}{2}(1\pm\frac{(a_0-a_1)+(b_0-b_1)\langle\hat{H}\hat{X}\hat{H}\rangle+(c_0-c_1)\langle\hat{H}\hat{Y}\hat{H}\rangle+(d_0-d_1)\langle\hat{H}\hat{Z}\hat{H}\rangle}{2\gamma^2-1})&\\
		&=\frac{1}{2}(1\pm\frac{(a_0-a_1)+(b_0-b_1)\langle\hat{Z}\rangle-(c_0-c_1)\langle\hat{Y}\rangle+(d_0-d_1)\langle\hat{X}\rangle}{2\gamma^2-1})&\\
		&=\frac{1}{2}(1\pm(\delta_a^{'} +\delta_d^{'} \cos2\alpha)),&
		\label{eq:pb}
	\end{aligned}
	\eeq
	where $\delta_a^{'}$ and $\delta_d^{'}$ in the Eq. (\ref{eq:pa}) and (\ref{eq:pb}) are the key parameters. It is obvious that the expressions in the Eq. (\ref{eq:pa}) and (\ref{eq:pb}) reduce to the ideal results  $p_{a\pm}=\frac{1}{2}(1\pm\cos2\alpha)$ and
	$p_{b\pm}=\frac{1}{2}(1\pm\sin2\alpha)$, respectively, when $e_1=\infty$, and $e_2=\infty$.
	
	\begin{figure}[htbp]
		\centering
		\includegraphics[width=16cm]{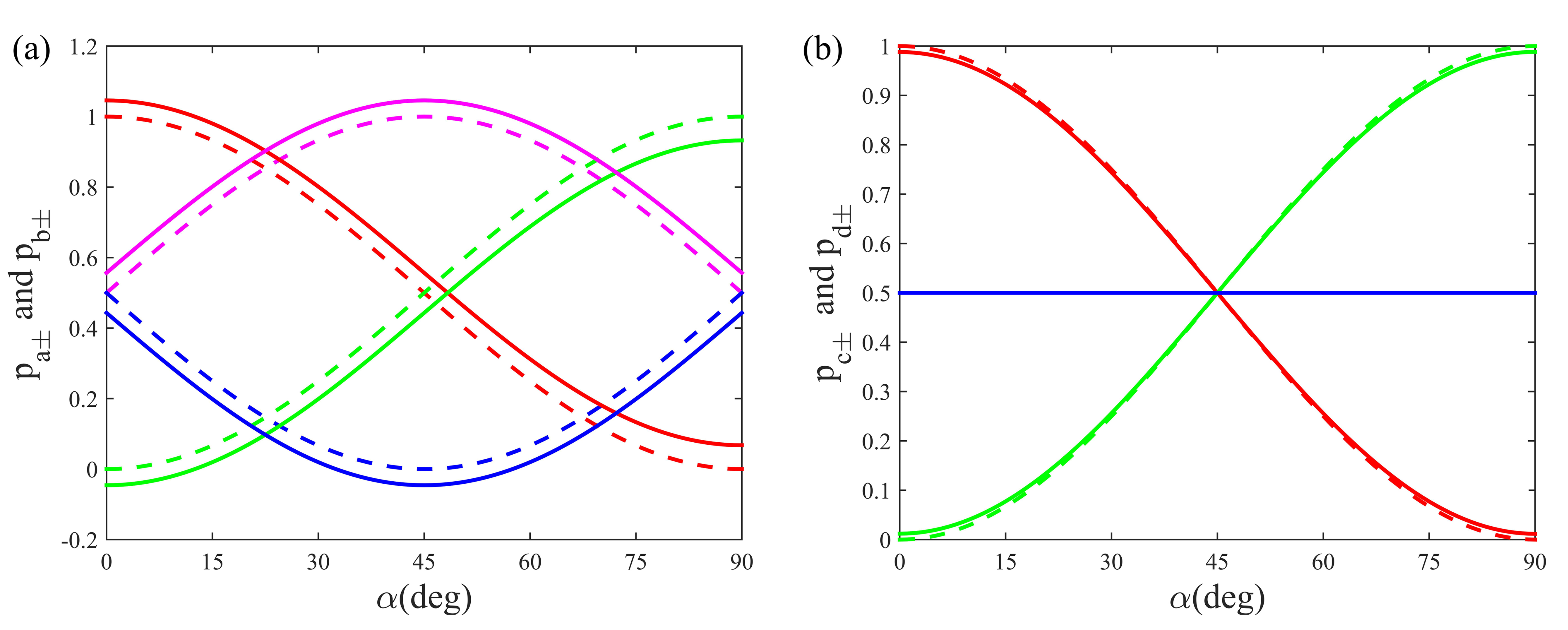}
		\caption{ The system state is  $\ket{\Phi}_{s}=\cos\alpha\ket{0}_{pol}+\sin\alpha\ket{1}_{pol}$. (a)The probabilities $p_{a\pm}$ and $p_{b\pm}$. (b) The probabilities $p_{c\pm}$ and $p_{d\pm}$. Dashed line: ideal theory, smooth line: theory with corrected imperfection in PBS.}
		\label{Fig.P}
	\end{figure}
	
	In the following, all dashed lines are the results with the perfect PBSs, while the smooth lines are the theoretical results with corrected imperfection in PBSs, $e_1=e_2=50$.
	
\begin{figure}[htb]
		\centering
		\includegraphics[width=10cm]{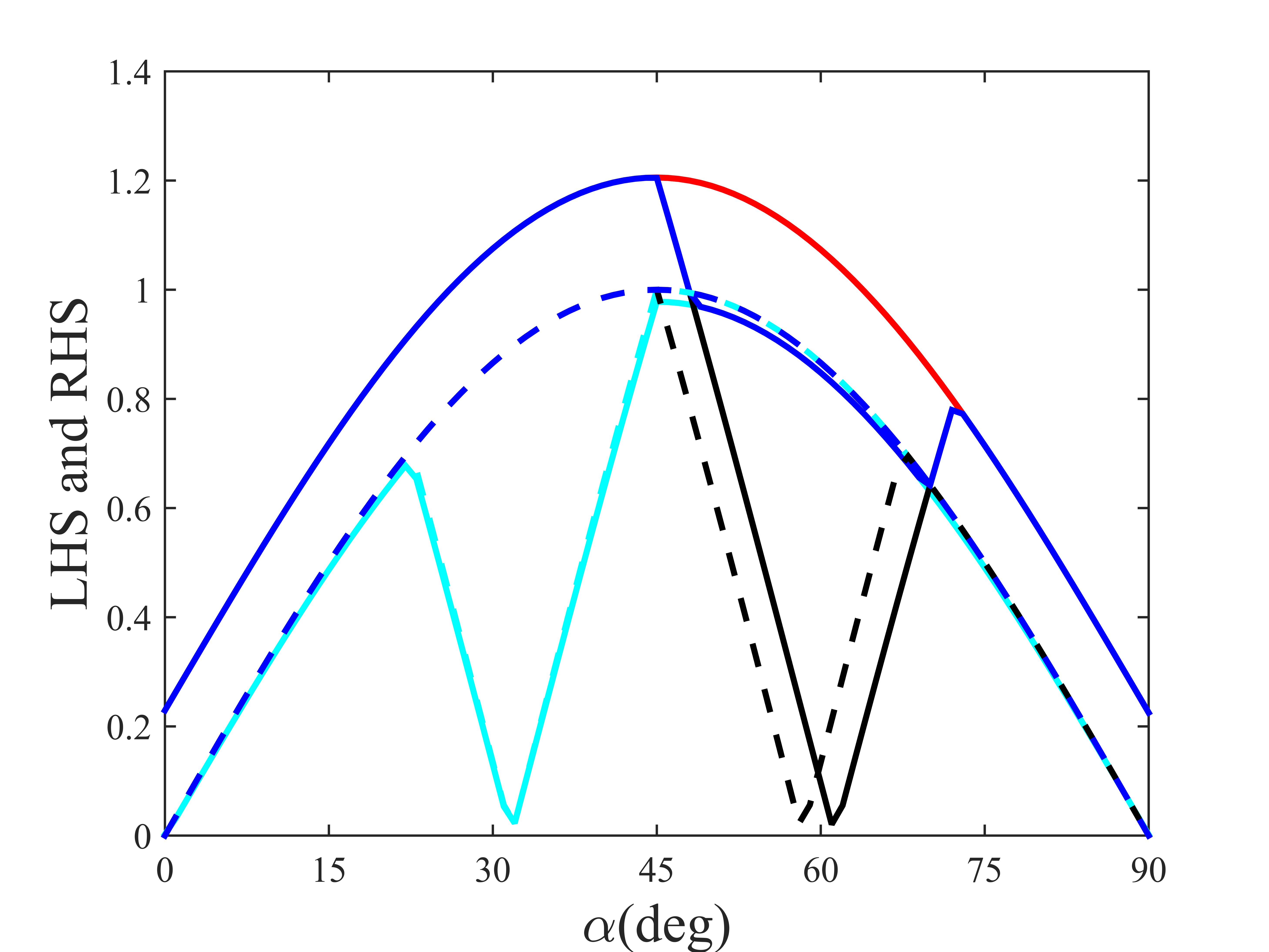}
		\caption{The results of LHS and RHS. The system state is  $\ket{\Phi}_{s}=\cos\alpha\ket{0}_{pol}+\sin\alpha\ket{1}_{pol}$. The angle $\theta$ in the measurement strength $\cos2\theta$ is set to $0^\circ$.  Dashed line: ideal theory, smooth line: theory with corrected imperfection in PBS.}
		\label{fig:LRHS}
	\end{figure}

	In the Fig. \ref{Fig.P}(a), $p_{a+}$, $p_{a-}$, $p_{b+}$ and $p_{b-}$ are plotted by red, green, purple, and blue lines, respectively. We note that $p_{a+}$ and $p_{b+}$ are greater than 1 for some input states, meanwhile the corresponding $p_{a-}$ and $p_{b-}$ are less than 0, due to the system imperfection in the experiment. While the impact is not so severe for the general measurements $\hat{C}$ and $\hat{D}$, whose probabilities are direct physical observations. The probabilities $p_{c+}$, $p_{c-}$, $p_{d+}$ and $p_{d-}$ are plotted by red, green, purple, and blue lines in the Fig. \ref{Fig.P}(b), respectively, with the measurement strength $\cos2\theta=1$. It is obvious that the $p_{c+}$ and $p_{c-}$ have a very slight difference in the ideal and unideal cases, they are symmetrical with respect to line $\alpha=45^{\circ}$.  Moreover, $p_{d+}$ and $p_{d-}$ are always 0.5. So the main influence to the LHS and RHS are the probabilities $p_{a+}$, $p_{a-}$, $p_{b+}$ and $p_{b-}$. By direct calculation we have
	\beq
	\begin{aligned}
		\xi_{G1} &=||p_{a+}-p_{b+}|+|p_{a-}-p_{b-}|-|p_{c+}-p_{d+}|-|p_{c-}-p_{d-}||,&\\
		\xi_{G2} &=||p_{a+}-p_{b-}|+|p_{a-}-p_{b+}|-|p_{c+}-p_{d-}|-|p_{c-}-p_{d+}||,
	\end{aligned}
	\eeq
	and $\xi_{G,max}=\max\{\xi_{G1},\xi_{G2}\}$, see Fig. \ref{fig:LRHS}, where the red, cyan, black, and blue lines correspond to the results $E_A+D_B$(LHS), $\xi_{G1}$, $\xi_{G2}$, $\xi_{G,max}$ (RHS). It is clear
	why the experimental results of LHS and RHS do not coincide for some system states.
	
	\section{Experimental results for the circularly polarized system qubit}
	
\begin{figure}[htb]
		\centering
		\includegraphics[width=8cm]{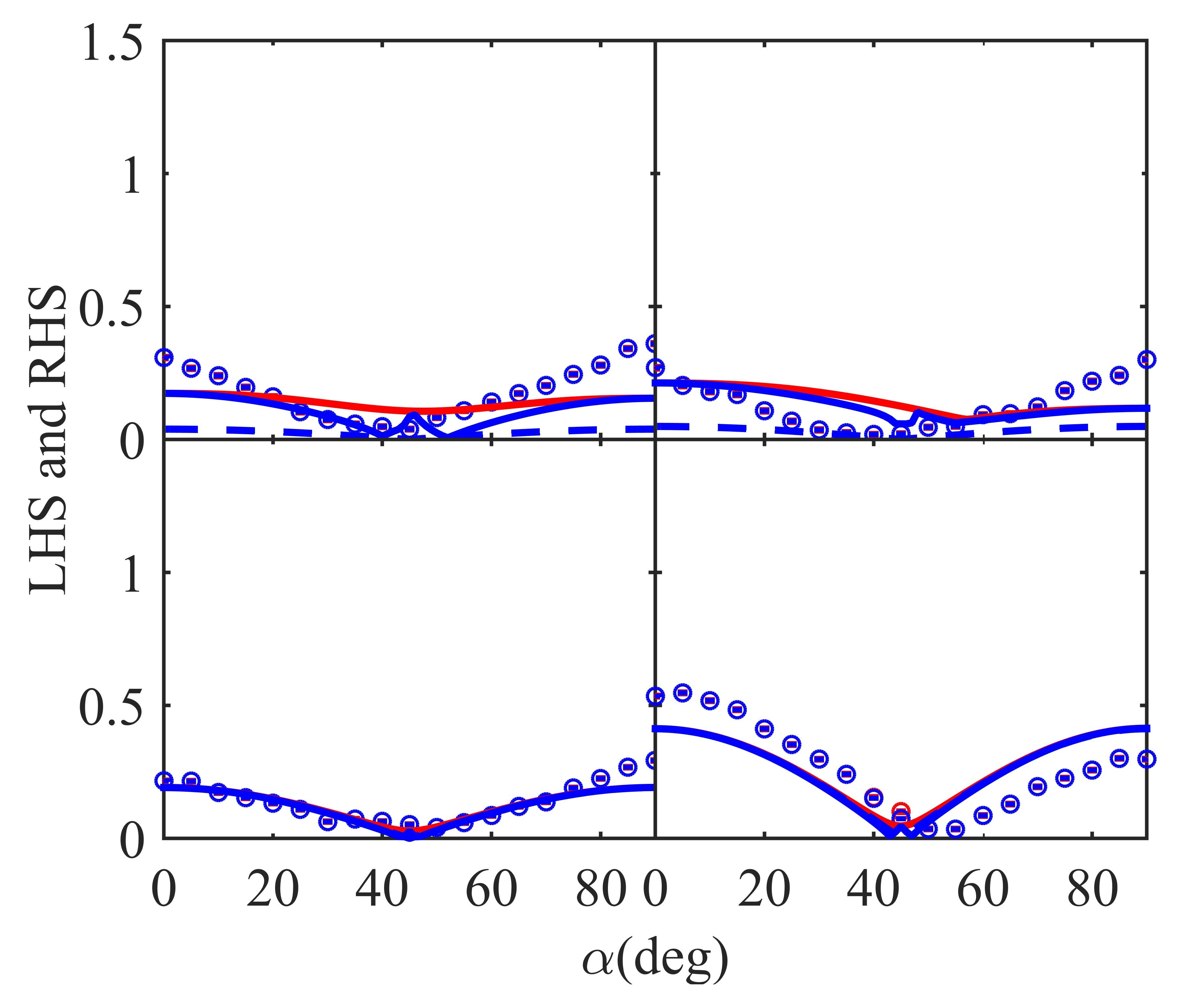}
		\caption{The experimetal results for circularly polarization states. (1) LHS and RHS. (2) Error and disturbance.The system state is  $\ket{\Phi}_{s}=\cos\alpha\ket{0}_{pol}+i\sin\alpha\ket{1}_{pol}$. From (a) to (d), the angle $\theta$ in the measurement strength $\cos2\theta$ is set to $0^\circ$, $9^\circ$, $18^\circ$, and $27^\circ$, respectively. Dashed line: ideal theory, smooth line: theory with corrected imperfection in PBS,  circles: experimental results. The error bars stand for one standard deviation, assuming poissonian statistics.}
		\label{fig:RL}
	\end{figure}

	For the circularly polarized system qubit, we also set the angle $\theta$ in the measurement strength $\cos2\theta$ to $0^\circ$, $9^\circ$, $18^\circ$, and $27^\circ$, respectively. In each measurement strength, we scan the angle $\alpha$ in the system state $\ket{\Phi}_{s}=\cos\alpha\ket{0}_{pol}+i\sin\alpha\ket{1}_{pol}$ from $0^\circ$ to $90^\circ$. All the experimental results of LHS (red circles) and RHS (blue circles) are shown in the Fig. \ref{fig:RL},  the former is always greater than or equal to the latter.
	
\bibliography{main}

\begin{thebibliography}{38}%
\makeatletter
\providecommand \@ifxundefined [1]{%
 \@ifx{#1\undefined}
}%
\providecommand \@ifnum [1]{%
 \ifnum #1\expandafter \@firstoftwo
 \else \expandafter \@secondoftwo
 \fi
}%
\providecommand \@ifx [1]{%
 \ifx #1\expandafter \@firstoftwo
 \else \expandafter \@secondoftwo
 \fi
}%
\providecommand \natexlab [1]{#1}%
\providecommand \enquote  [1]{``#1''}%
\providecommand \bibnamefont  [1]{#1}%
\providecommand \bibfnamefont [1]{#1}%
\providecommand \citenamefont [1]{#1}%
\providecommand \href@noop [0]{\@secondoftwo}%
\providecommand \href [0]{\begingroup \@sanitize@url \@href}%
\providecommand \@href[1]{\@@startlink{#1}\@@href}%
\providecommand \@@href[1]{\endgroup#1\@@endlink}%
\providecommand \@sanitize@url [0]{\catcode `\\12\catcode `\$12\catcode
  `\&12\catcode `\#12\catcode `\^12\catcode `\_12\catcode `\%12\relax}%
\providecommand \@@startlink[1]{}%
\providecommand \@@endlink[0]{}%
\providecommand \url  [0]{\begingroup\@sanitize@url \@url }%
\providecommand \@url [1]{\endgroup\@href {#1}{\urlprefix }}%
\providecommand \urlprefix  [0]{URL }%
\providecommand \Eprint [0]{\href }%
\providecommand \doibase [0]{http://dx.doi.org/}%
\providecommand \selectlanguage [0]{\@gobble}%
\providecommand \bibinfo  [0]{\@secondoftwo}%
\providecommand \bibfield  [0]{\@secondoftwo}%
\providecommand \translation [1]{[#1]}%
\providecommand \BibitemOpen [0]{}%
\providecommand \bibitemStop [0]{}%
\providecommand \bibitemNoStop [0]{.\EOS\space}%
\providecommand \EOS [0]{\spacefactor3000\relax}%
\providecommand \BibitemShut  [1]{\csname bibitem#1\endcsname}%
\let\auto@bib@innerbib\@empty
\bibitem [{\citenamefont {Busch}\ \emph
  {et~al.}(2014{\natexlab{a}})\citenamefont {Busch}, \citenamefont {Lahti},\
  and\ \citenamefont {Werner}}]{RevModPhys.86.1261}%
  \BibitemOpen
  \bibfield  {author} {\bibinfo {author} {\bibfnamefont {P.}~\bibnamefont
  {Busch}}, \bibinfo {author} {\bibfnamefont {P.}~\bibnamefont {Lahti}}, \ and\
  \bibinfo {author} {\bibfnamefont {R.~F.}\ \bibnamefont {Werner}},\ }\href
  {\doibase 10.1103/RevModPhys.86.1261} {\bibfield  {journal} {\bibinfo
  {journal} {Rev. Mod. Phys.}\ }\textbf {\bibinfo {volume} {86}},\ \bibinfo
  {pages} {1261} (\bibinfo {year} {2014}{\natexlab{a}})}\BibitemShut {NoStop}%
\bibitem [{\citenamefont {Coles}\ \emph {et~al.}(2017)\citenamefont {Coles},
  \citenamefont {Berta}, \citenamefont {Tomamichel},\ and\ \citenamefont
  {Wehner}}]{RevModPhys.89.015002}%
  \BibitemOpen
  \bibfield  {author} {\bibinfo {author} {\bibfnamefont {P.~J.}\ \bibnamefont
  {Coles}}, \bibinfo {author} {\bibfnamefont {M.}~\bibnamefont {Berta}},
  \bibinfo {author} {\bibfnamefont {M.}~\bibnamefont {Tomamichel}}, \ and\
  \bibinfo {author} {\bibfnamefont {S.}~\bibnamefont {Wehner}},\ }\href
  {\doibase 10.1103/RevModPhys.89.015002} {\bibfield  {journal} {\bibinfo
  {journal} {Rev. Mod. Phys.}\ }\textbf {\bibinfo {volume} {89}},\ \bibinfo
  {pages} {015002} (\bibinfo {year} {2017})}\BibitemShut {NoStop}%
\bibitem [{\citenamefont {Heisenberg}(1927)}]{Heisenberg1927}%
  \BibitemOpen
  \bibfield  {author} {\bibinfo {author} {\bibfnamefont {W.}~\bibnamefont
  {Heisenberg}},\ }\href {\doibase 10.1007/BF01397280} {\bibfield  {journal}
  {\bibinfo  {journal} {Z. Phys.}\ }\textbf {\bibinfo {volume} {43}},\ \bibinfo
  {pages} {172} (\bibinfo {year} {1927})}\BibitemShut {NoStop}%
\bibitem [{\citenamefont {Kennard}(1927)}]{Kennard1927}%
  \BibitemOpen
  \bibfield  {author} {\bibinfo {author} {\bibfnamefont {E.~H.}\ \bibnamefont
  {Kennard}},\ }\href {\doibase 10.1007/BF01391200} {\bibfield  {journal}
  {\bibinfo  {journal} {Z. Phys.}\ }\textbf {\bibinfo {volume} {44}},\ \bibinfo
  {pages} {326} (\bibinfo {year} {1927})}\BibitemShut {NoStop}%
\bibitem [{\citenamefont {Robertson}(1929)}]{robertson1929uncertainty}%
  \BibitemOpen
  \bibfield  {author} {\bibinfo {author} {\bibfnamefont {H.~P.}\ \bibnamefont
  {Robertson}},\ }\href {\doibase 10.1103/PhysRev.34.163} {\bibfield  {journal}
  {\bibinfo  {journal} {Phys. Rev.}\ }\textbf {\bibinfo {volume} {34}},\
  \bibinfo {pages} {163} (\bibinfo {year} {1929})}\BibitemShut {NoStop}%
\bibitem [{\citenamefont {Scully}\ \emph {et~al.}(1991)\citenamefont {Scully},
  \citenamefont {Englert},\ and\ \citenamefont {Walther}}]{Scully1991nature}%
  \BibitemOpen
  \bibfield  {author} {\bibinfo {author} {\bibfnamefont {M.~O.}\ \bibnamefont
  {Scully}}, \bibinfo {author} {\bibfnamefont {B.-G.}\ \bibnamefont {Englert}},
  \ and\ \bibinfo {author} {\bibfnamefont {H.}~\bibnamefont {Walther}},\ }\href
  {http://dx.doi.org/10.1038/351111a0} {\bibfield  {journal} {\bibinfo
  {journal} {Nature (London)}\ }\textbf {\bibinfo {volume} {351}},\ \bibinfo
  {pages} {111} (\bibinfo {year} {1991})}\BibitemShut {NoStop}%
\bibitem [{\citenamefont {Storey}\ \emph {et~al.}(1994)\citenamefont {Storey},
  \citenamefont {Tan}, \citenamefont {Collett},\ and\ \citenamefont
  {Walls}}]{Storey1994nature}%
  \BibitemOpen
  \bibfield  {author} {\bibinfo {author} {\bibfnamefont {P.}~\bibnamefont
  {Storey}}, \bibinfo {author} {\bibfnamefont {S.}~\bibnamefont {Tan}},
  \bibinfo {author} {\bibfnamefont {M.}~\bibnamefont {Collett}}, \ and\
  \bibinfo {author} {\bibfnamefont {D.}~\bibnamefont {Walls}},\ }\href
  {http://dx.doi.org/10.1038/367626a0} {\bibfield  {journal} {\bibinfo
  {journal} {Nature (London)}\ }\textbf {\bibinfo {volume} {367}},\ \bibinfo
  {pages} {626} (\bibinfo {year} {1994})}\BibitemShut {NoStop}%
\bibitem [{\citenamefont {Wiseman}\ and\ \citenamefont
  {Harrison}(1995)}]{Wiseman1995nature}%
  \BibitemOpen
  \bibfield  {author} {\bibinfo {author} {\bibfnamefont {H.}~\bibnamefont
  {Wiseman}}\ and\ \bibinfo {author} {\bibfnamefont {F.}~\bibnamefont
  {Harrison}},\ }\href {http://dx.doi.org/10.1038/377584a0} {\bibfield
  {journal} {\bibinfo  {journal} {Nature (London)}\ }\textbf {\bibinfo {volume}
  {377}},\ \bibinfo {pages} {584} (\bibinfo {year} {1995})}\BibitemShut
  {NoStop}%
\bibitem [{\citenamefont {Erhart}\ \emph {et~al.}(2012)\citenamefont {Erhart},
  \citenamefont {Sponar}, \citenamefont {Sulyok}, \citenamefont {Badurek},
  \citenamefont {Ozawa},\ and\ \citenamefont
  {Hasegawa}}]{erhart2012experimental}%
  \BibitemOpen
  \bibfield  {author} {\bibinfo {author} {\bibfnamefont {J.}~\bibnamefont
  {Erhart}}, \bibinfo {author} {\bibfnamefont {S.}~\bibnamefont {Sponar}},
  \bibinfo {author} {\bibfnamefont {G.}~\bibnamefont {Sulyok}}, \bibinfo
  {author} {\bibfnamefont {G.}~\bibnamefont {Badurek}}, \bibinfo {author}
  {\bibfnamefont {M.}~\bibnamefont {Ozawa}}, \ and\ \bibinfo {author}
  {\bibfnamefont {Y.}~\bibnamefont {Hasegawa}},\ }\href {\doibase
  10.1038/nphys2194} {\bibfield  {journal} {\bibinfo  {journal} {Nat. Phys.}\
  }\textbf {\bibinfo {volume} {8}},\ \bibinfo {pages} {185} (\bibinfo {year}
  {2012})}\BibitemShut {NoStop}%
\bibitem [{\citenamefont {Rozema}\ \emph {et~al.}(2012)\citenamefont {Rozema},
  \citenamefont {Darabi}, \citenamefont {Mahler}, \citenamefont {Hayat},
  \citenamefont {Soudagar},\ and\ \citenamefont
  {Steinberg}}]{PhysRevLett.109.100404}%
  \BibitemOpen
  \bibfield  {author} {\bibinfo {author} {\bibfnamefont {L.~A.}\ \bibnamefont
  {Rozema}}, \bibinfo {author} {\bibfnamefont {A.}~\bibnamefont {Darabi}},
  \bibinfo {author} {\bibfnamefont {D.~H.}\ \bibnamefont {Mahler}}, \bibinfo
  {author} {\bibfnamefont {A.}~\bibnamefont {Hayat}}, \bibinfo {author}
  {\bibfnamefont {Y.}~\bibnamefont {Soudagar}}, \ and\ \bibinfo {author}
  {\bibfnamefont {A.~M.}\ \bibnamefont {Steinberg}},\ }\href {\doibase
  10.1103/PhysRevLett.109.100404} {\bibfield  {journal} {\bibinfo  {journal}
  {Phys. Rev. Lett.}\ }\textbf {\bibinfo {volume} {109}},\ \bibinfo {pages}
  {100404} (\bibinfo {year} {2012})}\BibitemShut {NoStop}%
\bibitem [{\citenamefont {Baek}\ \emph {et~al.}(2013)\citenamefont {Baek},
  \citenamefont {Kaneda}, \citenamefont {Ozawa},\ and\ \citenamefont
  {Edamatsu}}]{baek2013experimental}%
  \BibitemOpen
  \bibfield  {author} {\bibinfo {author} {\bibfnamefont {S.-Y.}\ \bibnamefont
  {Baek}}, \bibinfo {author} {\bibfnamefont {F.}~\bibnamefont {Kaneda}},
  \bibinfo {author} {\bibfnamefont {M.}~\bibnamefont {Ozawa}}, \ and\ \bibinfo
  {author} {\bibfnamefont {K.}~\bibnamefont {Edamatsu}},\ }\href
  {https://doi.org/10.1038/srep02221} {\bibfield  {journal} {\bibinfo
  {journal} {Sci. Rep.}\ }\textbf {\bibinfo {volume} {3}},\ \bibinfo {pages}
  {2221} (\bibinfo {year} {2013})}\BibitemShut {NoStop}%
\bibitem [{\citenamefont {Sulyok}\ \emph {et~al.}(2013)\citenamefont {Sulyok},
  \citenamefont {Sponar}, \citenamefont {Erhart}, \citenamefont {Badurek},
  \citenamefont {Ozawa},\ and\ \citenamefont {Hasegawa}}]{PhysRevA.88.022110}%
  \BibitemOpen
  \bibfield  {author} {\bibinfo {author} {\bibfnamefont {G.}~\bibnamefont
  {Sulyok}}, \bibinfo {author} {\bibfnamefont {S.}~\bibnamefont {Sponar}},
  \bibinfo {author} {\bibfnamefont {J.}~\bibnamefont {Erhart}}, \bibinfo
  {author} {\bibfnamefont {G.}~\bibnamefont {Badurek}}, \bibinfo {author}
  {\bibfnamefont {M.}~\bibnamefont {Ozawa}}, \ and\ \bibinfo {author}
  {\bibfnamefont {Y.}~\bibnamefont {Hasegawa}},\ }\href {\doibase
  10.1103/PhysRevA.88.022110} {\bibfield  {journal} {\bibinfo  {journal} {Phys.
  Rev. A}\ }\textbf {\bibinfo {volume} {88}},\ \bibinfo {pages} {022110}
  (\bibinfo {year} {2013})}\BibitemShut {NoStop}%
\bibitem [{\citenamefont {Kaneda}\ \emph {et~al.}(2014)\citenamefont {Kaneda},
  \citenamefont {Baek}, \citenamefont {Ozawa},\ and\ \citenamefont
  {Edamatsu}}]{PhysRevLett.112.020402}%
  \BibitemOpen
  \bibfield  {author} {\bibinfo {author} {\bibfnamefont {F.}~\bibnamefont
  {Kaneda}}, \bibinfo {author} {\bibfnamefont {S.-Y.}\ \bibnamefont {Baek}},
  \bibinfo {author} {\bibfnamefont {M.}~\bibnamefont {Ozawa}}, \ and\ \bibinfo
  {author} {\bibfnamefont {K.}~\bibnamefont {Edamatsu}},\ }\href {\doibase
  10.1103/PhysRevLett.112.020402} {\bibfield  {journal} {\bibinfo  {journal}
  {Phys. Rev. Lett.}\ }\textbf {\bibinfo {volume} {112}},\ \bibinfo {pages}
  {020402} (\bibinfo {year} {2014})}\BibitemShut {NoStop}%
\bibitem [{\citenamefont {Ringbauer}\ \emph {et~al.}(2014)\citenamefont
  {Ringbauer}, \citenamefont {Biggerstaff}, \citenamefont {Broome},
  \citenamefont {Fedrizzi}, \citenamefont {Branciard},\ and\ \citenamefont
  {White}}]{PhysRevLett.112.020401}%
  \BibitemOpen
  \bibfield  {author} {\bibinfo {author} {\bibfnamefont {M.}~\bibnamefont
  {Ringbauer}}, \bibinfo {author} {\bibfnamefont {D.~N.}\ \bibnamefont
  {Biggerstaff}}, \bibinfo {author} {\bibfnamefont {M.~A.}\ \bibnamefont
  {Broome}}, \bibinfo {author} {\bibfnamefont {A.}~\bibnamefont {Fedrizzi}},
  \bibinfo {author} {\bibfnamefont {C.}~\bibnamefont {Branciard}}, \ and\
  \bibinfo {author} {\bibfnamefont {A.~G.}\ \bibnamefont {White}},\ }\href
  {\doibase 10.1103/PhysRevLett.112.020401} {\bibfield  {journal} {\bibinfo
  {journal} {Phys. Rev. Lett.}\ }\textbf {\bibinfo {volume} {112}},\ \bibinfo
  {pages} {020401} (\bibinfo {year} {2014})}\BibitemShut {NoStop}%
\bibitem [{\citenamefont {Ozawa}(2003)}]{Ozawa03}%
  \BibitemOpen
  \bibfield  {author} {\bibinfo {author} {\bibfnamefont {M.}~\bibnamefont
  {Ozawa}},\ }\href {\doibase 10.1103/PhysRevA.67.042105} {\bibfield  {journal}
  {\bibinfo  {journal} {Phys. Rev. A}\ }\textbf {\bibinfo {volume} {67}},\
  \bibinfo {pages} {042105} (\bibinfo {year} {2003})}\BibitemShut {NoStop}%
\bibitem [{\citenamefont {Ozawa}(2004{\natexlab{a}})}]{ozawa2004uncertainty}%
  \BibitemOpen
  \bibfield  {author} {\bibinfo {author} {\bibfnamefont {M.}~\bibnamefont
  {Ozawa}},\ }\href {\doibase 10.1016/j.aop.2003.12.012} {\bibfield  {journal}
  {\bibinfo  {journal} {Ann. Phys.}\ }\textbf {\bibinfo {volume} {311}},\
  \bibinfo {pages} {350} (\bibinfo {year} {2004}{\natexlab{a}})}\BibitemShut
  {NoStop}%
\bibitem [{\citenamefont {Ozawa}(2004{\natexlab{b}})}]{OZAWA2004367}%
  \BibitemOpen
  \bibfield  {author} {\bibinfo {author} {\bibfnamefont {M.}~\bibnamefont
  {Ozawa}},\ }\href {\doibase https://doi.org/10.1016/j.physleta.2003.12.001}
  {\bibfield  {journal} {\bibinfo  {journal} {Phys. Lett. A}\ }\textbf
  {\bibinfo {volume} {320}},\ \bibinfo {pages} {367 } (\bibinfo {year}
  {2004}{\natexlab{b}})}\BibitemShut {NoStop}%
\bibitem [{\citenamefont {Branciard}(2013)}]{branciard2013error}%
  \BibitemOpen
  \bibfield  {author} {\bibinfo {author} {\bibfnamefont {C.}~\bibnamefont
  {Branciard}},\ }\href {\doibase 10.1073/pnas.1219331110} {\bibfield
  {journal} {\bibinfo  {journal} {Proc. Natl. Acad. Sci. U.S.A}\ }\textbf
  {\bibinfo {volume} {110}},\ \bibinfo {pages} {6742} (\bibinfo {year}
  {2013})}\BibitemShut {NoStop}%
\bibitem [{\citenamefont {Di~Lorenzo}(2013)}]{PhysRevLett.110.120403}%
  \BibitemOpen
  \bibfield  {author} {\bibinfo {author} {\bibfnamefont {A.}~\bibnamefont
  {Di~Lorenzo}},\ }\href {\doibase 10.1103/PhysRevLett.110.120403} {\bibfield
  {journal} {\bibinfo  {journal} {Phys. Rev. Lett.}\ }\textbf {\bibinfo
  {volume} {110}},\ \bibinfo {pages} {120403} (\bibinfo {year}
  {2013})}\BibitemShut {NoStop}%
\bibitem [{\citenamefont {Weston}\ \emph {et~al.}(2013)\citenamefont {Weston},
  \citenamefont {Hall}, \citenamefont {Palsson}, \citenamefont {Wiseman},\ and\
  \citenamefont {Pryde}}]{PhysRevLett.110.220402}%
  \BibitemOpen
  \bibfield  {author} {\bibinfo {author} {\bibfnamefont {M.~M.}\ \bibnamefont
  {Weston}}, \bibinfo {author} {\bibfnamefont {M.~J.~W.}\ \bibnamefont {Hall}},
  \bibinfo {author} {\bibfnamefont {M.~S.}\ \bibnamefont {Palsson}}, \bibinfo
  {author} {\bibfnamefont {H.~M.}\ \bibnamefont {Wiseman}}, \ and\ \bibinfo
  {author} {\bibfnamefont {G.~J.}\ \bibnamefont {Pryde}},\ }\href {\doibase
  10.1103/PhysRevLett.110.220402} {\bibfield  {journal} {\bibinfo  {journal}
  {Phys. Rev. Lett.}\ }\textbf {\bibinfo {volume} {110}},\ \bibinfo {pages}
  {220402} (\bibinfo {year} {2013})}\BibitemShut {NoStop}%
\bibitem [{\citenamefont {Busch}\ \emph {et~al.}(2013)\citenamefont {Busch},
  \citenamefont {Lahti},\ and\ \citenamefont {Werner}}]{Werner1}%
  \BibitemOpen
  \bibfield  {author} {\bibinfo {author} {\bibfnamefont {P.}~\bibnamefont
  {Busch}}, \bibinfo {author} {\bibfnamefont {P.}~\bibnamefont {Lahti}}, \ and\
  \bibinfo {author} {\bibfnamefont {R.~F.}\ \bibnamefont {Werner}},\ }\href
  {\doibase 10.1103/PhysRevLett.111.160405} {\bibfield  {journal} {\bibinfo
  {journal} {Phys. Rev. Lett.}\ }\textbf {\bibinfo {volume} {111}},\ \bibinfo
  {pages} {160405} (\bibinfo {year} {2013})}\BibitemShut {NoStop}%
\bibitem [{\citenamefont {Branciard}(2014)}]{PhysRevA.89.022124}%
  \BibitemOpen
  \bibfield  {author} {\bibinfo {author} {\bibfnamefont {C.}~\bibnamefont
  {Branciard}},\ }\href {\doibase 10.1103/PhysRevA.89.022124} {\bibfield
  {journal} {\bibinfo  {journal} {Phys. Rev. A}\ }\textbf {\bibinfo {volume}
  {89}},\ \bibinfo {pages} {022124} (\bibinfo {year} {2014})}\BibitemShut
  {NoStop}%
\bibitem [{\citenamefont {Buscemi}\ \emph
  {et~al.}(2014{\natexlab{a}})\citenamefont {Buscemi}, \citenamefont {Hall},
  \citenamefont {Ozawa},\ and\ \citenamefont {Wilde}}]{PhysRevLett.112.050401}%
  \BibitemOpen
  \bibfield  {author} {\bibinfo {author} {\bibfnamefont {F.}~\bibnamefont
  {Buscemi}}, \bibinfo {author} {\bibfnamefont {M.~J.~W.}\ \bibnamefont
  {Hall}}, \bibinfo {author} {\bibfnamefont {M.}~\bibnamefont {Ozawa}}, \ and\
  \bibinfo {author} {\bibfnamefont {M.~M.}\ \bibnamefont {Wilde}},\ }\href
  {\doibase 10.1103/PhysRevLett.112.050401} {\bibfield  {journal} {\bibinfo
  {journal} {Phys. Rev. Lett.}\ }\textbf {\bibinfo {volume} {112}},\ \bibinfo
  {pages} {050401} (\bibinfo {year} {2014}{\natexlab{a}})}\BibitemShut
  {NoStop}%
\bibitem [{\citenamefont {Busch}\ \emph
  {et~al.}(2014{\natexlab{b}})\citenamefont {Busch}, \citenamefont {Lahti},\
  and\ \citenamefont {Werner}}]{Werner2}%
  \BibitemOpen
  \bibfield  {author} {\bibinfo {author} {\bibfnamefont {P.}~\bibnamefont
  {Busch}}, \bibinfo {author} {\bibfnamefont {P.}~\bibnamefont {Lahti}}, \ and\
  \bibinfo {author} {\bibfnamefont {R.~F.}\ \bibnamefont {Werner}},\ }\href
  {\doibase 10.1063/1.4871444} {\bibfield  {journal} {\bibinfo  {journal} {J.
  Math. Phys.}\ }\textbf {\bibinfo {volume} {55}},\ \bibinfo {pages} {042111}
  (\bibinfo {year} {2014}{\natexlab{b}})}\BibitemShut {NoStop}%
\bibitem [{\citenamefont {Busch}\ \emph
  {et~al.}(2014{\natexlab{c}})\citenamefont {Busch}, \citenamefont {Lahti},\
  and\ \citenamefont {Werner}}]{Werner3}%
  \BibitemOpen
  \bibfield  {author} {\bibinfo {author} {\bibfnamefont {P.}~\bibnamefont
  {Busch}}, \bibinfo {author} {\bibfnamefont {P.}~\bibnamefont {Lahti}}, \ and\
  \bibinfo {author} {\bibfnamefont {R.~F.}\ \bibnamefont {Werner}},\ }\href
  {\doibase 10.1103/PhysRevA.89.012129} {\bibfield  {journal} {\bibinfo
  {journal} {Phys. Rev. A}\ }\textbf {\bibinfo {volume} {89}},\ \bibinfo
  {pages} {012129} (\bibinfo {year} {2014}{\natexlab{c}})}\BibitemShut
  {NoStop}%
\bibitem [{\citenamefont {Lu}\ \emph {et~al.}(2014)\citenamefont {Lu},
  \citenamefont {Yu}, \citenamefont {Fujikawa},\ and\ \citenamefont
  {Oh}}]{PhysRevA.90.042113}%
  \BibitemOpen
  \bibfield  {author} {\bibinfo {author} {\bibfnamefont {X.-M.}\ \bibnamefont
  {Lu}}, \bibinfo {author} {\bibfnamefont {S.}~\bibnamefont {Yu}}, \bibinfo
  {author} {\bibfnamefont {K.}~\bibnamefont {Fujikawa}}, \ and\ \bibinfo
  {author} {\bibfnamefont {C.~H.}\ \bibnamefont {Oh}},\ }\href {\doibase
  10.1103/PhysRevA.90.042113} {\bibfield  {journal} {\bibinfo  {journal} {Phys.
  Rev. A}\ }\textbf {\bibinfo {volume} {90}},\ \bibinfo {pages} {042113}
  (\bibinfo {year} {2014})}\BibitemShut {NoStop}%
\bibitem [{\citenamefont {Dressel}\ and\ \citenamefont
  {Nori}(2014)}]{PhysRevA.89.022106}%
  \BibitemOpen
  \bibfield  {author} {\bibinfo {author} {\bibfnamefont {J.}~\bibnamefont
  {Dressel}}\ and\ \bibinfo {author} {\bibfnamefont {F.}~\bibnamefont {Nori}},\
  }\href {\doibase 10.1103/PhysRevA.89.022106} {\bibfield  {journal} {\bibinfo
  {journal} {Phys. Rev. A}\ }\textbf {\bibinfo {volume} {89}},\ \bibinfo
  {pages} {022106} (\bibinfo {year} {2014})}\BibitemShut {NoStop}%
\bibitem [{\citenamefont {Korzekwa}\ \emph {et~al.}(2014)\citenamefont
  {Korzekwa}, \citenamefont {Jennings},\ and\ \citenamefont
  {Rudolph}}]{PhysRevA.89.052108}%
  \BibitemOpen
  \bibfield  {author} {\bibinfo {author} {\bibfnamefont {K.}~\bibnamefont
  {Korzekwa}}, \bibinfo {author} {\bibfnamefont {D.}~\bibnamefont {Jennings}},
  \ and\ \bibinfo {author} {\bibfnamefont {T.}~\bibnamefont {Rudolph}},\ }\href
  {\doibase 10.1103/PhysRevA.89.052108} {\bibfield  {journal} {\bibinfo
  {journal} {Phys. Rev. A}\ }\textbf {\bibinfo {volume} {89}},\ \bibinfo
  {pages} {052108} (\bibinfo {year} {2014})}\BibitemShut {NoStop}%
\bibitem [{\citenamefont {Sulyok}\ \emph {et~al.}(2015)\citenamefont {Sulyok},
  \citenamefont {Sponar}, \citenamefont {Demirel}, \citenamefont {Buscemi},
  \citenamefont {Hall}, \citenamefont {Ozawa},\ and\ \citenamefont
  {Hasegawa}}]{PhysRevLett.115.030401}%
  \BibitemOpen
  \bibfield  {author} {\bibinfo {author} {\bibfnamefont {G.}~\bibnamefont
  {Sulyok}}, \bibinfo {author} {\bibfnamefont {S.}~\bibnamefont {Sponar}},
  \bibinfo {author} {\bibfnamefont {B.}~\bibnamefont {Demirel}}, \bibinfo
  {author} {\bibfnamefont {F.}~\bibnamefont {Buscemi}}, \bibinfo {author}
  {\bibfnamefont {M.~J.~W.}\ \bibnamefont {Hall}}, \bibinfo {author}
  {\bibfnamefont {M.}~\bibnamefont {Ozawa}}, \ and\ \bibinfo {author}
  {\bibfnamefont {Y.}~\bibnamefont {Hasegawa}},\ }\href {\doibase
  10.1103/PhysRevLett.115.030401} {\bibfield  {journal} {\bibinfo  {journal}
  {Phys. Rev. Lett.}\ }\textbf {\bibinfo {volume} {115}},\ \bibinfo {pages}
  {030401} (\bibinfo {year} {2015})}\BibitemShut {NoStop}%
\bibitem [{\citenamefont {Busch}\ and\ \citenamefont
  {Stevens}(2015)}]{PhysRevLett.114.070402}%
  \BibitemOpen
  \bibfield  {author} {\bibinfo {author} {\bibfnamefont {P.}~\bibnamefont
  {Busch}}\ and\ \bibinfo {author} {\bibfnamefont {N.}~\bibnamefont
  {Stevens}},\ }\href {\doibase 10.1103/PhysRevLett.114.070402} {\bibfield
  {journal} {\bibinfo  {journal} {Phys. Rev. Lett.}\ }\textbf {\bibinfo
  {volume} {114}},\ \bibinfo {pages} {070402} (\bibinfo {year}
  {2015})}\BibitemShut {NoStop}%
\bibitem [{\citenamefont {Demirel}\ \emph {et~al.}(2016)\citenamefont
  {Demirel}, \citenamefont {Sponar}, \citenamefont {Sulyok}, \citenamefont
  {Ozawa},\ and\ \citenamefont {Hasegawa}}]{PhysRevLett.117.140402}%
  \BibitemOpen
  \bibfield  {author} {\bibinfo {author} {\bibfnamefont {B.}~\bibnamefont
  {Demirel}}, \bibinfo {author} {\bibfnamefont {S.}~\bibnamefont {Sponar}},
  \bibinfo {author} {\bibfnamefont {G.}~\bibnamefont {Sulyok}}, \bibinfo
  {author} {\bibfnamefont {M.}~\bibnamefont {Ozawa}}, \ and\ \bibinfo {author}
  {\bibfnamefont {Y.}~\bibnamefont {Hasegawa}},\ }\href {\doibase
  10.1103/PhysRevLett.117.140402} {\bibfield  {journal} {\bibinfo  {journal}
  {Phys. Rev. Lett.}\ }\textbf {\bibinfo {volume} {117}},\ \bibinfo {pages}
  {140402} (\bibinfo {year} {2016})}\BibitemShut {NoStop}%
\bibitem [{\citenamefont {Ma}\ \emph {et~al.}(2016)\citenamefont {Ma},
  \citenamefont {Ma}, \citenamefont {Wang}, \citenamefont {Chen}, \citenamefont
  {Liu}, \citenamefont {Kong}, \citenamefont {Li}, \citenamefont {Peng},
  \citenamefont {Shi}, \citenamefont {Shi}, \citenamefont {Fei},\ and\
  \citenamefont {Du}}]{EXPW1}%
  \BibitemOpen
  \bibfield  {author} {\bibinfo {author} {\bibfnamefont {W.}~\bibnamefont
  {Ma}}, \bibinfo {author} {\bibfnamefont {Z.}~\bibnamefont {Ma}}, \bibinfo
  {author} {\bibfnamefont {H.}~\bibnamefont {Wang}}, \bibinfo {author}
  {\bibfnamefont {Z.}~\bibnamefont {Chen}}, \bibinfo {author} {\bibfnamefont
  {Y.}~\bibnamefont {Liu}}, \bibinfo {author} {\bibfnamefont {F.}~\bibnamefont
  {Kong}}, \bibinfo {author} {\bibfnamefont {Z.}~\bibnamefont {Li}}, \bibinfo
  {author} {\bibfnamefont {X.}~\bibnamefont {Peng}}, \bibinfo {author}
  {\bibfnamefont {M.}~\bibnamefont {Shi}}, \bibinfo {author} {\bibfnamefont
  {F.}~\bibnamefont {Shi}}, \bibinfo {author} {\bibfnamefont {S.-M.}\
  \bibnamefont {Fei}}, \ and\ \bibinfo {author} {\bibfnamefont
  {J.}~\bibnamefont {Du}},\ }\href {\doibase 10.1103/PhysRevLett.116.160405}
  {\bibfield  {journal} {\bibinfo  {journal} {Phys. Rev. Lett.}\ }\textbf
  {\bibinfo {volume} {116}},\ \bibinfo {pages} {160405} (\bibinfo {year}
  {2016})}\BibitemShut {NoStop}%
\bibitem [{\citenamefont {Zhou}\ \emph {et~al.}(2016)\citenamefont {Zhou},
  \citenamefont {Yan}, \citenamefont {Gong}, \citenamefont {Ma}, \citenamefont
  {He}, \citenamefont {Xiong}, \citenamefont {Chen}, \citenamefont {Yang},
  \citenamefont {Feng},\ and\ \citenamefont {Vedral}}]{EXPW2}%
  \BibitemOpen
  \bibfield  {author} {\bibinfo {author} {\bibfnamefont {F.}~\bibnamefont
  {Zhou}}, \bibinfo {author} {\bibfnamefont {L.}~\bibnamefont {Yan}}, \bibinfo
  {author} {\bibfnamefont {S.}~\bibnamefont {Gong}}, \bibinfo {author}
  {\bibfnamefont {Z.}~\bibnamefont {Ma}}, \bibinfo {author} {\bibfnamefont
  {J.}~\bibnamefont {He}}, \bibinfo {author} {\bibfnamefont {T.}~\bibnamefont
  {Xiong}}, \bibinfo {author} {\bibfnamefont {L.}~\bibnamefont {Chen}},
  \bibinfo {author} {\bibfnamefont {W.}~\bibnamefont {Yang}}, \bibinfo {author}
  {\bibfnamefont {M.}~\bibnamefont {Feng}}, \ and\ \bibinfo {author}
  {\bibfnamefont {V.}~\bibnamefont {Vedral}},\ }\href {\doibase
  10.1126/sciadv.1600578} {\bibfield  {journal} {\bibinfo  {journal} {Sci.
  Adv.}\ }\textbf {\bibinfo {volume} {2}},\ \bibinfo {pages} {e1600578}
  (\bibinfo {year} {2016})}\BibitemShut {NoStop}%
\bibitem [{\citenamefont {Sulyok}\ and\ \citenamefont
  {Sponar}(2017)}]{PhysRevA.96.022137}%
  \BibitemOpen
  \bibfield  {author} {\bibinfo {author} {\bibfnamefont {G.}~\bibnamefont
  {Sulyok}}\ and\ \bibinfo {author} {\bibfnamefont {S.}~\bibnamefont
  {Sponar}},\ }\href {\doibase 10.1103/PhysRevA.96.022137} {\bibfield
  {journal} {\bibinfo  {journal} {Phys. Rev. A}\ }\textbf {\bibinfo {volume}
  {96}},\ \bibinfo {pages} {022137} (\bibinfo {year} {2017})}\BibitemShut
  {NoStop}%
\bibitem [{\citenamefont {Barchielli}\ \emph {et~al.}(2018)\citenamefont
  {Barchielli}, \citenamefont {Gregoratti},\ and\ \citenamefont
  {Toigo}}]{Barchielli2018}%
  \BibitemOpen
  \bibfield  {author} {\bibinfo {author} {\bibfnamefont {A.}~\bibnamefont
  {Barchielli}}, \bibinfo {author} {\bibfnamefont {M.}~\bibnamefont
  {Gregoratti}}, \ and\ \bibinfo {author} {\bibfnamefont {A.}~\bibnamefont
  {Toigo}},\ }\href {\doibase 10.1007/s00220-017-3075-7} {\bibfield  {journal}
  {\bibinfo  {journal} {Commun. Math. Phys.}\ }\textbf {\bibinfo {volume}
  {357}},\ \bibinfo {pages} {1253} (\bibinfo {year} {2018})}\BibitemShut
  {NoStop}%
\bibitem [{\citenamefont {Buscemi}\ \emph
  {et~al.}(2014{\natexlab{b}})\citenamefont {Buscemi}, \citenamefont {Hall},
  \citenamefont {Ozawa},\ and\ \citenamefont {Wilde}}]{buscemipaying}%
  \BibitemOpen
  \bibfield  {author} {\bibinfo {author} {\bibfnamefont {F.}~\bibnamefont
  {Buscemi}}, \bibinfo {author} {\bibfnamefont {M.~J.}\ \bibnamefont {Hall}},
  \bibinfo {author} {\bibfnamefont {M.}~\bibnamefont {Ozawa}}, \ and\ \bibinfo
  {author} {\bibfnamefont {M.~M.}\ \bibnamefont {Wilde}},\ }\href@noop {}
  {\bibfield  {journal} {\bibinfo  {journal} {Asia-Pacic Workshop on Quantum
  Information Sciences}\ } (\bibinfo {year} {National Cheng Kung University,
  Taiwan, 2014}{\natexlab{b}})}\BibitemShut {NoStop}%
\bibitem [{\citenamefont {Wilmott}(2011)}]{qutrits}%
  \BibitemOpen
  \bibfield  {author} {\bibinfo {author} {\bibfnamefont {C.~M.}\ \bibnamefont
  {Wilmott}},\ }\href {\doibase 10.1142/S0219749911008143} {\bibfield
  {journal} {\bibinfo  {journal} {Int. J. Quantum Inf.}\ }\textbf {\bibinfo
  {volume} {09}},\ \bibinfo {pages} {1511} (\bibinfo {year}
  {2011})}\BibitemShut {NoStop}%
\bibitem [{\citenamefont {Lund}\ and\ \citenamefont {Wiseman}(2010)}]{Lund}%
  \BibitemOpen
  \bibfield  {author} {\bibinfo {author} {\bibfnamefont {A.~P.}\ \bibnamefont
  {Lund}}\ and\ \bibinfo {author} {\bibfnamefont {H.~M.}\ \bibnamefont
  {Wiseman}},\ }\href {http://stacks.iop.org/1367-2630/12/i=9/a=093011}
  {\bibfield  {journal} {\bibinfo  {journal} {New J. Phys.}\ }\textbf {\bibinfo
  {volume} {12}},\ \bibinfo {pages} {093011} (\bibinfo {year}
  {2010})}\BibitemShut {NoStop}%
\end{thebibliography}%
\end{document}